# Understanding the environmental impacts of virgin aggregates: critical literature review and primary comprehensive Life Cycle Assessments

Anne de Bortoli[1,2,3,4]*

**1** CIRAIG, École Polytechnique de Montréal, P.O. Box 6079, Montréal, Québec, H3C 3A7, Canada

**2** Direction technique, Eurovia Management, 18 Place de l'Europe, 92500, Rueil-Malmaison, France

**3** Centre Technique Amériques, Eurovia Canada Inc., 3705 Place Java #210, Brossard, QC J4Y 0E4, Canada

**4** LVMT, Ecole des Ponts ParisTech, Cité Descartes, 6-8 Avenue Blaise Pascal, 77420 Champs-sur-Marne, France

**\*** Corresponding author; e-mail: anne.debortoli@polymtl.ca

ABSTRACT

Despite the ever-growing massive consumption of aggregates, knowledge about their environmental footprint is limited. My literature review on virgin aggregate Life Cycle Assessments (LCA) highlighted many shortcomings, such as low-quality inputs and fragmented system boundaries, and estimated that gravel consumption is responsible for 0.17 to 1.8% of the global carbon footprint. I thus developed comprehensive LCAs, based on field data collected in Quebec's quarries producing annually 7 million tons of aggregates, representing different types of rocks, productions (mobile, fixed), and energies consumed, using ecoinvent 3.7 and TRACI characterization method. Results show that the often-forgotten blasting and machinery are major contributors to several impact categories, along with diesel consumption. The link between the nature of the rock and the aggregate's environmental impacts is demonstrated for the first time: the harder it is, the more explosive it requires, thus increasing the impacts. Moreover, the more abrasive the rock is, the faster it wears out machinery, generating higher maintenance that increases human and ecosystem toxicities. A pronounced sensitivity of the impacts to the electricity mix is also shown based on a scenario analysis carried on Europe, China, and different Canadian and American regions. Additionally, aggregate transportation to the consumer, modeled with tailored inventories, can more than double the impact of the aggregate at quarry's gate, with strong regional variability. In a near future, I call for considering consistent system boundaries in aggregate LCA, refining blasting, energy consumption, machinery manufacturing and maintenance, as well as customizing truck transportation models, for more reliable aggregate LCAs.

# 1. Introduction

## 1.1 Context

Despite aggregates being the most consumed material worldwide, with an ever-growing demand (Bendixen et al., 2021; Miatto et al., 2017), knowledge about their life cycle environmental footprint is currently limited and uncertain. Miatto et al. estimated that each human being consumed almost 5 tons of non-metallic resources on average in 2010 – three quarter of them being gravel and sand -, accounting for a total of 34 billion metric tons (Gt) extracted annually worldwide (Miatto et al. 2017). This massive production is responsible for serious environmental impacts. Near-site qualitative impacts mainly include landscape deterioration, noise, dust, potential sedimentation, and pollution of water bodies. Land use and land use change (LULUC) related to aggregate extraction (Langer and Arbogast, 2002), as well as the potential environmental consequences of LULUC such as loss of habitats and ecosystem erosion, greenhouse gas (GHG) releases, and deterioration of air quality (Foley et al., 2005), also jeopardize the environment . But some environmental impacts obviously appear at larger scales than at the quarry scale, over the entire life cycle of the aggregate. Namely, some impacts occur during the quarry preparation (i.e., removal of the top soil layer), the usage and end-of-life (EoL) of the aggregates, and the various production stages including extraction, transportation, crushing, and sieving. For instance, during these activities, the use of vehicles, building machines, and other production equipment such as crushing units, consume energy whose supply chains are complex, emitting GHGs (Ghanbari et al., 2018) and other pollutants in several areas worldwide.

To quantify comprehensively the different kinds of environmental impacts of aggregate production and consumption – *i.e.,* at local and global scales, over their entire life cycle, and including the supply chain of all the consumptions related to this production -, the most adequate method is life cycle assessment (LCA), as the only method to finely assess on the life cycle and from a multicriteria point of view quantitative impacts of products. LCA is a method normalized by ISO 14040 and 14044 standards (International Organization for Standardization, 2006a, 2006b) that assesses quantitatively and on different environmental dimensions the impacts of a system over its life cycle, from the extraction of raw material to the end-of-life. Many aggregate LCAs have already been conducted based on real production sites or site archetypes, in different contexts and on different system boundaries, especially on recycled aggregates from concrete (Zhang et al., 2019). Nevertheless, virgin gravel and sand are the most important kinds of non-mineral resources extracted in the world, respectively representing 41 and 31% of the global production (Miatto et al., 2017), and recycled aggregates only represented an estimated 9% of the aggregate market in 2022 (Fact.MR, 2023). The limited usage of recycled aggregates is partly linked to the quality reduction of this material compared to virgin aggregates (Zhang et al., 2019). I thus propose to examine further LCAs published on virgin gravels - that I will call indifferently crushed stones or aggregates in the rest of this article – to understand the state of advancement of aggregate LCA modeling, its potential shortcomings, and their potential consequences on the understanding of aggregate environmental impacts.

## 1.2 *Critical literature review*

### 1.2.1. Method

I performed this literature review on virgin aggregate LCAs between Springs 2022 and 2023, mainly through sciencedirect.com and Google Scholar, using the association of key words "LCA" (or its developed form) and "aggregates" or "gravel". I analyzed the main elements from the first three steps of performing an LCA according to ISO 14040, namely goal and scopes specifics, life cycle inventories (LCI), as well as characterization methods. In an LCA, the LCI is the list and quantities of input and output flows needed to deliver the functional unit, the functional unit being an elementary quantity of the product or service analyzed. The quality of this step is crucial for the reliability of the LCA results. Thus, I also reported characteristics giving information on the quality of the LCA based on the Pedigree Matrix quality dimensions: geography, data collection period, number of sites assessed, nature of the activity, etc. A few virgin aggregate LCAs have not been reported because they were too far from being ISO 14040 and 14044-compliant (International Organization for Standardization, 2006a, 2006b). This is the case of studies from Mladenovic et al. whose LCI of natural aggregates has been "gathered from available datasets" (no further information) (Mladenovič et al., 2015), from Marinković et al. who only reported in their LCA the consumption of diesel and its emissions (total of 9 elementary flows) for excavating river gravel and loading/unloading a ship with an elevator (Marinković et al., 2008), and from Simion et al., reported in the literature review of the study by Martinez-Arguelles et al., but that gives too limited information (Martinez-Arguelles et al., 2019; Simion et al., 2013). Some studies have also been mistakenly reported as primary LCAs on virgin aggregates by Dias et al. (Dias et al., 2022) and were, of course, excluded from our literature review. First, some of these studies use or report secondary LCIs, like Tošić et al. using the data from Marinković et al. (Marinković et al., 2010; Tošić et al., 2015), Fraj and Idir using the LCIs from ecoinvent v2.2 (Fraj and Idir, 2017; Kellenberger et al., 2007), Kurda et al. using the data from Marinković et al. and Braga et al. (Braga et al., 2017; Marinković et al., 2010), and Park et al. (Park et al., 2019). Second, some studies are not LCAs, but life cycle cost analyses (LCCAs) or mechanical studies on aggregates (Kulekci et al., 2021; Ohemeng and Ekolu, 2020; Tam, 2008).

*1.2.2. Analysis*

Following the method described previously, nineteen LCAs or LCIs on natural aggregates have been published with enough transparency to be reported in this literature review. Table 1 presents the related publications and the main characteristics of the LCA models. The studies have been published from 2001 to 2022. Only nine studies provide reproducible LCIs, by giving access to their complete models (de Bortoli et al., 2022), or their primary inventory as well as the information on the background LCIs used (Federal Highway Administration, 2020; Rosado et al., 2017; Ananth and Mundada, 2017; Lesage and Samson, 2016; Jullien et al., 2012; Kellenberger et al., 2007; Stripple, 2001). Three studies do not even provide their background LCIs, unless the information has been missed (Estanqueiro et al., 2018; Faleschini et al., 2016; Ghanbari et al., 2018). When a life cycle impact assessment (LCIA) is performed, different characterization methods are used, from the oldest methods such as Ecoindicator 99 (Colangelo et al., 2018; Estanqueiro et al., 2018), to the most common methods such as CML (Braga et al., 2017; Estanqueiro et al., 2018; Faleschini et al., 2016; Jullien et al., 2012; Pradhan et al., 2019) and Impact 2002+ (Estanqueiro et al., 2018; Gan et al., 2016; Martinez-Arguelles et al., 2019; Rosado et al., 2017), or the most up-to-date ReCiPe and ImpactWORLD+ (IW+) methods (de Bortoli et al., 2022).

Table 1 LCIs and LCAs of natural aggregates: publications, main model characteristics and reproducibility

| Authors | Publication date | Publisher | Reproducible LCIs | Background LCIs | LCIA method |
|---|---|---|---|---|---|
| *Stripple* | 2001 | IVL | * | Elementary flows | N/A |
| *Kellenberger et al.* | 2007 | EI v2 | * | EI v3 | N/A |
| *Korre and Durucan* | 2009 | WRAP | | Elementary flows | CML baseline |

| Jullien et al. | 2012 | Elsevier | * | Elementary flows | Modified CML 2001 |
| Lesage and Samson | 2016 | Springer/EI v3 | * | EI v3 | N/A |
| Estanqueiro et al. | 2016 | Taylor & Francis | | ? | CML 2000, Ecoindicator 99, CED |
| Hossain et al. | 2016 | Elsevier | | China Light and Power, Chinese Life Cycle Database, European reference Life Cycle Database | Impact 2002+ |
| Faleschini et al. | 2016 | Elsevier | | ? | CML 2002 |
| Gan et al. | 2016 | Elsevier | | Elementary flows | Impact 2002+ |
| Ananth et al. | 2017 | EI v3 | * | EI v3 | N/A |
| Braga et al. | 2017 | Elsevier | | EI v3 | CML baseline |
| Rosado et al. | 2017 | Elsevier | * | EI v3, EU & DK input-output database, US LCI | Impact 2002+ |
| Silva et al. | 2017 | SRI/EI v3 | * | EI v3 | ILCD LCIA |
| Ghanbari et al. | 2018 | Springer | | ? | ? |
| Colangelo et al. | 2018 | Elsevier | | EI v3, literature, etc. | Ecoindicator 99 |
| Pradhan et al. | 2019 | Elsevier | | EI v3.01 | CML baseline |
| Martinez-Arguelles et al. | 2019 | SAGE | | EI v3.0, US LCI | Impact 2002+ |
| Federal Highway Administration | 2020 | Federal LCA Commons | * | Elementary flows | N/A |
| de Bortoli et al. | 2022 | MPDI | * | EI v2.2 | IW+/ReCiPe |

Table 2 describes the scope of each LCI. Unclear information has been noted in brackets. First, only six studies consider the land use and land use consumption (LULUC) from the quarries. Then, only two studies clearly include the impacts due to clearing the quarry before starting the operations. Within the rock extraction stage, almost all the studies consider the excavation, loading and handling, but only nine clearly consider the consumption of blasting agent to dynamite hard rocks, and only six studies consider the emissions from blasting. The LCI supplied in ecoinvent since the version 2 are heterogeneous: the generic "gravel production" processes do not considered blasting, while they may be the most used in LCA modes, while the "limestone, unprocessed" process does (Kellenberger et al., 2007). Silva et al. precise that particulate matters (PM) emitted due to blasting are not accounted for in their model (Silva et al., 2018). Gan et al. only consider $CO_2$, $SO_2$ and NOx within blasting emissions, while Jullien et al. consider 9 atmospheric emissions including 5 pollutants and excluding PM. Rosado et al. as well as Silva et al. use the "*blasting*" process from ecoinvent, accounting for the consumption of explosive and its related blasting emissions, that include the same 5 types of pollutants plus 3 size categories of PM. Thus, the blasting process from ecoinvent is more comprehensive in terms of flows. Moreover, when comparing explosive consumption to dynamite one kilogram of stone, I calculated a factor of 5 between the Indian and EI limestone LCIs (resp. 7.09 and 7.73 $10^{-5}$ kg per kilogram of stone) and the Brazilian LCI by Silva et al. (3.7 $10^{-4}$ kg), while the explosive considered (Tovex) is the same. The Brazilian LCI by Rosado et al. indicates that 145 g of blasting agent is used per ton of gravel, i.e., an intermediary consumption of 1.45 $10^{-4}$ kg/kg. This indicates either a wide variability between sites and/or practices, or a large uncertain on these values. Korre and Durucan declare to account for the explosive, yet their inventory only includes energy flows. Jullien et al. consider on one of their sites the emissions from two of the three kinds of explosives used – fuel nitrate and emulsion. But the modelling of the blasting agent consumption is unclear, and at least does not consider the emissions to produce the explosives. Transportation of the rocks from quarry to crusher, as well as crushing and screening activities, are almost always accounted for. Nevertheless, when by truck, the studies mostly consider theoretical loads and fuel consumptions, and mostly forget empty return trips. Conversely, only four studies consider the amortization of some capital goods –conveyor when applicable, crushing unit, buildings when applicable, and infrastructure when applicable. Only the seminal LCIs in ecoinvent (Kellenberger et al., 2007) and its derivative dataset for Quebec (Lesage and Samson, 2016), as well as the Brazilian study by Silva et al. account for all of them. Surprisingly, the two first datasets consider metallic buildings, while no expert reported the presence of such buildings on the quarries in Quebec and France. About capital goods' amortization, only these three studies and the study by de Bortoli et al. consider the amortization of the crushing unit, modeled using the ecoinvent's generic process "*industrial machine, heavy, unspecified*", thus not accounting for the specific compounds of such machines. Only six studies clearly consider some machinery maintenance, mainly lubricating oil, and spare parts for building machines, through the accounting of steel and rubber flows. Specific metals and wearing parts are not represented. A few studies consider transportation from quarry to customer, but most of them account for generic distances instead of statistical distances, and model the impacts using the generic truck transportation processes from ecoinvent, while they are very far from being representative of transportation in the construction sector, with average loads around 6 tons for the 16-32 ton-type truck, instead of much higher tonnages in construction. Moreover, only two studies clearly consider the EoL of the quarry, including recultivation, restoration or final redevelopment of the site after shutdown. Finally, a few specificities have not been reported in Table 2: only Ananth et al. and de Bortoli et al. clearly include the consumption from lighting in the quarry, and only this latter team include administrative activities.

Besides the system boundaries considered, the exhaustivity of the flows considered in the LCIs is also important for the quality and thus the robustness of the assessments. Yet, several studies restrain their inventory to a few flows. This is for instance the case of the "production of aggregate" process in the Federal LCA Commons (Federal Highway Administration, 2020) and the study by Ghanbari et al. that only account for energy consumption flows and their related emissions (Federal Highway Administration, 2020; Ghanbari et al., 2018), or the study by Martinez-Arguelles et al that only accounts for diesel, lubricants, and transportation economic flows. Such approaches generate large uncertainties on numerous impact categories.

**Table 2 Scope of the LCI/LCAs: system boundaries**

| Publication | LULUC | Clea-ring | Extraction Explosive (kg/kg stone) | Extraction Blasting emissions | Extraction Excavating, loading, handling | Quarry to crusher transport | Crushing & screening | Capital goods' amortization Conveyor | Capital goods' amortization Crushing unit | Capital goods' amortization Buildings | Capital goods' amortization Infra-structure | Machinery Life-span | Machinery Maintenance | Down-stream transport | Quarry EoL |
|---|---|---|---|---|---|---|---|---|---|---|---|---|---|---|---|
| (Stripple, 2001) | | | | | * | * | * | | | | | N/A | | ? | |
| (Kellenberger et al., 2007) | * | | (*) | (*) | * | * | * | * | * | Steel | * | 25 | Wearing parts, oil | (*) | (*) |
| (Korre and Durucan, 2009) | | * | (*) | (*) | * | * | * | | | | | N/A | | | * |
| (Jullien et al., 2012) | | | (*) | | * | * | * | | | | | N/A | | | |
| (Lesage and Samson, 2016) | * | | | | * | * | * | * | * | Steel | * | 25 | Wearing parts, oil | (*) | (*) |
| (Estanqueiro et al., 2018) | | | * | ? | * | * | * | | | | | N/A | | * | |
| (Hossain et al., 2016) | | | | | * | * | * | | | | | N/A | | | |
| (Faleschini et al., 2016) | * | | | | * | * | * | | | | | N/A | | | |
| (Gan et al., 2016) | | | * | * | * | * | * | | | | | N/A | | * | |
| (Ananth and Mundada, 2017) | * | | * | | | | * | * | | | (*) | ? | Oil | (*) | |
| (Braga et al., 2017) | | | * | | * | * | * | | | | | N/A | | * | |
| (Rosado et al., 2017) | | | * | * | * | * | * | | | | | N/A | | * | |
| (Silva et al., 2018) | * | | * | * | * | * | * | * | * | * | * | 25 | Wearing parts, oil | ? | ? |
| (Ghanbari et al., 2018) | | | (N/A) | (N/A) | * | * | * | | | | | N/A | | | |
| (Colangelo et al., 2018) | * | ? | ? | ? | (*) | (*) | (*) | | | | | N/A | ? | ? | ? |
| (Pradhan et al., 2019) | | | * | | * | | * | | | | | N/A | | * | |
| (Martinez-Arguelles et al., 2019) | | | | | * | * | * | | | | | N/A | Oil | | |
| (Federal Highway Administration, 2020) | | ? | | | (*) | (*) | (*) | | | | | N/A | | ? | ? |
| (de Bortoli et al., 2022) | | | * | * | * | * | * | * | * | | | ? | Wearing parts, oil | * | * |

I now want to analyze the quality of the models published, based on the five dimensions of the Pedigree Matrix, where each dimensions get a score between 1 (best grade) and 5 (worst grade) quantifying the quality of the input data (Weidema, 1998; Weidema et al., 2013; Ciroth, 2013), Table 3 reports the geographic representativeness of the studies, the data collection period, as well as several elements related to the characteristics of aggregate production technologies: type and nature of rocks, mineral losses considered in the process, number of quarries analyzed and related annual production, as well as quarry areas. The studies are based on quarries mostly in Europe (9), then Asia (5), South America (3), and finally in North America (2). Only eight studies specifically report the period of data collection (Kellenberger et al., 2007; Jullien et al., 2012; Faleschini et al., 2016; Hossain et al., 2016; Lesage and Samson, 2016; Rosado et al., 2017; de Bortoli et al., 2022), while it is one of the information needed to assess the quality of an LCA with the Pedigree Matrix. The studies not reporting any period would automatically get the score of 5 in terms of temporal representativeness. The periods of data collection run from 1997 to 2016: the youngest data would thus get a score of 3 to represent 2023's conditions, while the other datasets would get a score of 4 or 5. Almost all the studies indicate the category of rock exploited (crushed or round, also called massive or loose rocks), but rarer are the studies specifying the nature of the rock. The mineral losses along the process of aggregate production are accounted for in half the studies, and range between 0 and 44%. Other studies do not mention any consideration of losses, even equal to zero. Two thirds of the studies give the number of quarries assessed – between one and sixteen, other studies would get a score of 5 in terms of completeness. For those specifying the number of sites studied, the production completeness is highly limited at the country scale, while studies often indicate national representativeness. For example, the study by Silva et al., published in ecoinvent, is tagged as "Brazilian", i.e. representative of Brazil aggregate production, while the 3 sites audited only represent 2.1% of the national production. The completeness score is thus of 3, while studies with one site audited get a score of 4. Thus, only five studies specify the quarry area – between 12 and 5810 kilo-square meters (ksm). Only the LCA by Kellenberger et al., reused by Lesage and Samson, and Silva et al. specifies resp. a 50- and 80-year operating time of the quarry. Nevertheless, all the studies that excluded LULUC from the scope of assessment are not concerned by this characteristic.

Table 3 Representativeness and quality of natural aggregate LCIs and LCAs

| Publication | Geography | Data collection period | Type of rock | Nature of rock | Mineral losses (%) | Number of quarries | Quarry area (ksm) | Production (t/y) |
|---|---|---|---|---|---|---|---|---|
| (Stripple, 2001) | Sweden | | Crushed | | 0 | 1 | | |
| (Kellenberger et al., 2007) | Switzerland | 1997-2001 | Round and crushed | | 4 | 4 | 12 | (1.6k-2M) |
| (Korre and Durucan, 2009) | United Kingdom | | Round and crushed | 7 types | 1-50 | | | |
| (Jullien et al., 2012) | France | 2004 | Round and crushed | Amphibolites, alluvial, Fz and Fy | | 3 | | 70k-180k |
| (Lesage and Samson, 2016) | Quebec | 1997-2001 | Crushed | | 4 | 4 | 12 | 1.6k-2M |
| (Estanqueiro et al., 2018) | Portugal | | Crushed | (Limestone) | 4 | | | |
| (Hossain et al., 2016) | China/Hong Kong | 2013 | Round and crushed | Undefined, river sand | | | | |
| (Faleschini et al., 2016) | Vicenza, Italy | 2010-2014 | (Crushed) | | | 1 | 100 | 56k |
| (Gan et al., 2016) | Hong Kong | | | | | (2) | | |
| (Ananth and Mundada, 2017) | North India | 2009-… | Crushed | (Granite and black trap stone) | | 1 | | |
| (Braga et al., 2017) | Portugal | | Round and crushed | Granite, limestone, undefined | | | | |
| (Rosado et al., 2017) | Sao Paulo, Brazil | 2015 | Crushed | Basalt | 0 | 1 | 860 | 450k |
| (Silva et al. 2018) | Sao Paulo, Brazil | 2015-2016 | Crushed | Granite | 0 | 3 | ? | ? |
| (Ghanbari et al., 2018) | Iran | | (Round) | (Marine gravel) | | (1) | | (480k) |
| (Colangelo et al., 2018) | Southern Italy | | | Sand & gravel | 8 | | | |
| (Pradhan et al., 2019) | India | | Crushed | Basalt | 0 | (1) | | (390k) |
| (Martinez-Arguelles et al., 2019) | Northern Colombia | | Crushed | Limestone | 44 | 1 | 5810 | 588k |
| (Federal Highway Administration, 2020) | United States | | | | | | | |
| (de Bortoli et al., 2022) | France | 2007-2016 | Round and crushed | | 0 | 16 | | |

As different characterization methods have been used in the LCAs studied (see Table 1), the environmental indicators assessed vary, as well as which phenomena they include for a same impact or damage category, and which part of the life cycle is considered due to system boundary discrepancies. This raised a problem on non-comparability of results. In Figure 1, I thus only present the results of the studies on the most popular indicator: the carbon footprint, also called Global Warming Potential (GWP), or contribution to climate change, and expressed in kilograms of $CO_2$ equivalent ($kgCO_2eq$). The impact assessment methods used in the different studies are still numerous but generate limited discrepancies in the impacts due to $CO_2$ being the main GHG emitted for the production of aggregates and having a stable characterization factor. The LCIs included in ecoinvent – namely those from Kellenberger et al., Lesage and Samson, Silva et al., Ananth et al. – are assessed in Figure 1 with the IPCC 2021 climate change method, GWP100, with a cut-off allocation. The impact of the production of natural aggregates from cradle to quarry gate are presented in magenta: they range from 1.42 to 35.58 $kgCO_2eq$ per ton (t) produced respectively in the study by Stripple and Martinez-Arguelles et al. The contribution of transportation from quarry to customer, when accounted, is presented in striped gray. It varies greatly. In absolute value, it ranges from 1.35 to 30.7 $kgCO_2eq/t$ resp. in Gan et al. Ananth et al., and in contribution, from 30% to 80%. Let's note that Braga et al. do not disaggregate their impact between production and downstream transportation as their study primarily focuses on concretes using different aggregates. In the end, from cradle to customer, the aggregates present a carbon footprint ranging from 3.67 to 38.2 kgCO2eq/t, resp. in Rosado et al. and Ananth et al. studies. This leads to a global carbon footprint of global aggregate production and delivery ranging between 93.9 and 974 Mt CO2eq, based on Miatto et al.' aggregate production estimate (2017), i.e., between 0.17 and 1.8% of global GHG emissions (Ritchie and Roser, 2022).

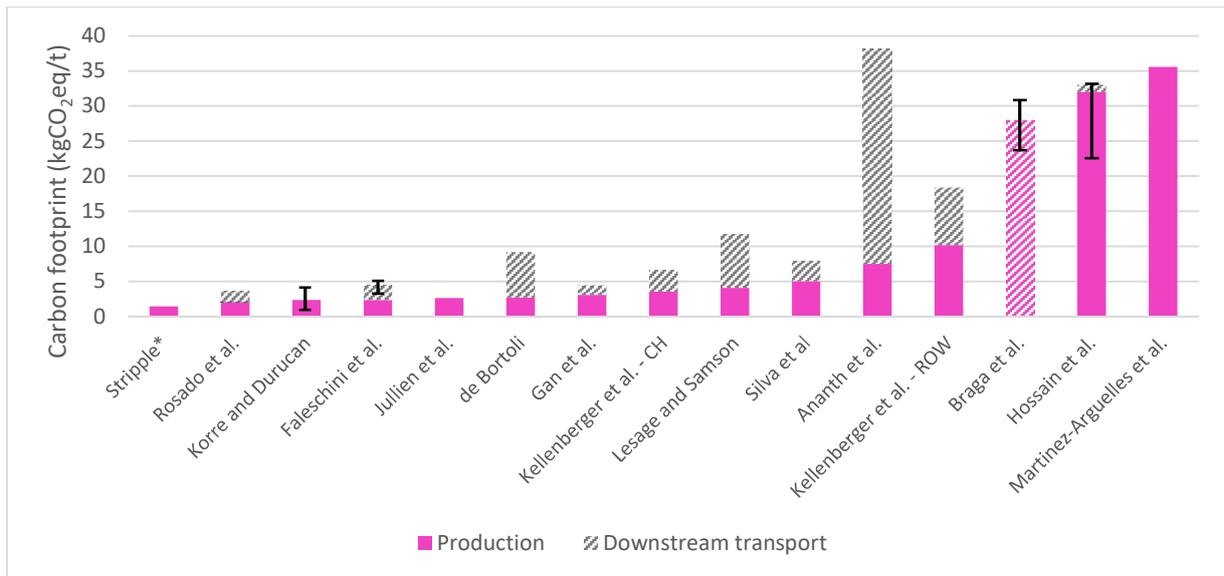

**Figure 1 Carbon footprint of natural aggregates in published LCAs, from cradle to quarry gate (magenta), due to transportation to customer (striped gray), or from cradle to customer (striped magenta). Interval of results is indicated in black.**

Let's also note that some of the studies in the previous tables do not appear on the graph: they do not provide the carbon footprint of natural aggregates, either because their main focus is not on aggregates production LCA – but relates to concrete (Colangelo et al., 2018; Estanqueiro et al., 2018; Pradhan et al., 2019) or to the national scale (Ghanbari et al., 2018) – or because they just include LCIs of elementary flows (Federal Highway Administration, 2020).

Finally, some of the studies also present the impact of recycled aggregates. It is interested to note that most studies present smaller production impacts for these aggregates than for virgin aggregate (Faleschini et al., 2016; Hossain et al., 2016; Martinez-Arguelles et al., 2019), nevertheless it is not systematic, and Gan et al found the opposite. This is due to the importance of transportation for these process, also emphasized by Gan et al. when showing that imported aggregates have a doubled carbon footprint compared to local aggregates. Round aggregates also seem to be less carbon-intense than crushed aggregates (e.g. in Hossain et al., 2016).

*1.2.3. Gaps and limitations of previous LCAs*

As a conclusion, this literature review on natural aggregate LCIs and LCAs highlights numerous shortcomings of the models published so far: low-quality input data, with weak technological, temporal, and geographical representativeness, small samples, obsolete and/or unrepresentative data, missing or unclear information, varied system boundaries, and lack of reproducibility for two thirds of the studies. Based on this literature review, the system boundaries always exclude the use stage and end-of-life of the aggregates themselves. This is probably due to the many possible usages of aggregates, mostly mixed with other materials, e.g. in concrete, in different environmental conditions, that generate different impacts. On the classical cradle-to-gate perimeter, the common exclusion of blasting consumption and/or emissions for hard rock exploitation is surprising - e.g. in the global and European ecoinvent models. The variability of explosive consumed to dynamite hard rocks also needs to be better understood, as a factor of 5 has been found between studies without technological reason. The exclusion of capital goods, or the consideration of generic industrial machine amortization as well as theoretical transportation distances and modeling also bring potential major errors in the results. Moreover, the sole consideration of energy consumptions in a few studies invalidate their multicriteria results. These multiple limits call for developing a comprehensive and up-to-date model, exploring different production processes, as well as types and nature of rocks, the interlink between these factors and the final environmental impact, and considering statistically representative transportation phases and capital goods' amortization and maintenance.

### 1.3 *Study's objectives, overview, and novelty*

To fill the gaps identified in section 1.2 and better understand the variability of inventories and environmental impacts related to the production of aggregates from massive rocks, I propose to perform comprehensive LCAs for different types of aggregates based on recent field data.

The main stages of the study are presented in Figure 2. The literature review (section 1.2) allowed to highlight some main shortcomings and discrepancies in previous virgin aggregate LCAs. It revealed the comprehensive system boundaries to consider in the goal and scope of a gravel LCA (section 2.1). It also showed "weak" LCIs used to assess the impact of aggregates, i.e., low quality and partial inputs to investigate further in a comprehensive LCA. A complementary pre-screening to capture the potential main contributors to the environmental impacts of virgin aggregates will be done (section 2.2.2) to validate the prioritization of data collections, data collection being time-consuming but key to high-quality assessments. Then, inventories will be developed accordingly in section 2.3 on: blasting (2.3.1), energy and water consumptions (2.3.2), dust emissions (2.3.3), facilities and machinery (2.3.4), and gravel downstream transportation (2.3.4). This inventory will be assessed to show results in section 3 on: gravels environmental hotspots (3.1), absolute values of the impact of different types of aggregates (3.2), and scenario analyses (3.3). The scenario analyses explore the variability of aggregate's impacts depending on the nature of rocks (3.3.1), the electricity mix (3.3.2), the nature of the installation – mobile versus fixed units (3.3.3), and transportation to the consumer (3.3.4) – distance, type of truck, tailored transportation modeling versus generic ecoinvent modeling.

The main novelties in this study will be first to consider a complete system perimeter and as exhaustive as possible flows, as well as to develop and use high quality input data - in terms of completeness and representativeness. Moreover, primary models specific to the consumption of explosive agents depending on the nature of rocks, capital goods construction/manufacturing and maintenance, as well as gravel transport by loaded and empty trucks, will be developed. In the end, the study aims to be the most advanced in terms of quality and therefore robustness of aggregate LCA results. A step further, this study aims to make the factors impacting the environmental performance of virgin aggregates in Quebec and in different areas of the world more understandable, by adopting an innovative fine disaggregation of the types of production of virgin aggregates. Finally, this model can serve as a basis for subsequent studies, because it is transparent and reproducible.

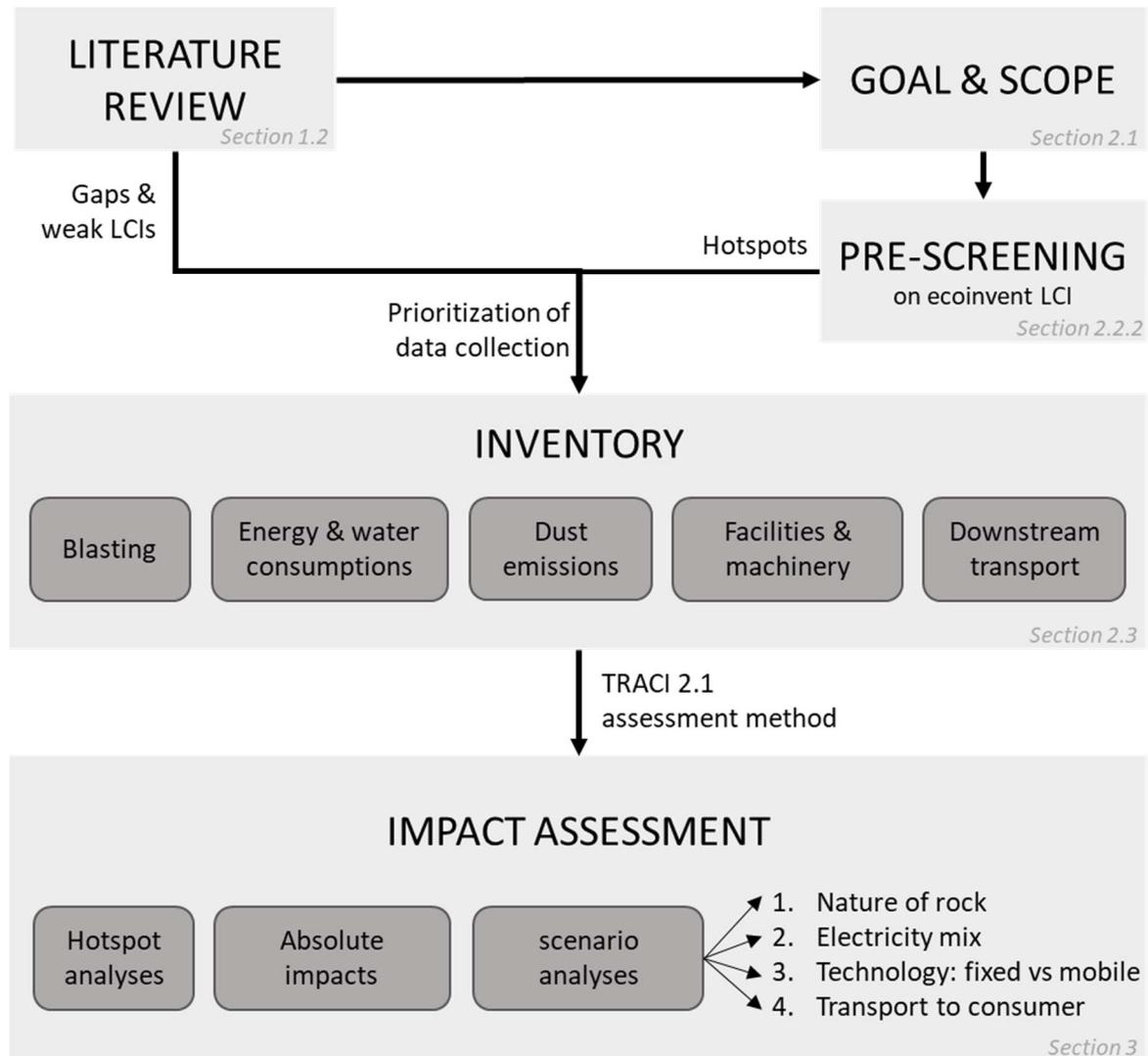

Figure 2 Flow chart: main stages of the study

# 2 Method

## *2.1. Goal and scope*

My goal is to carry out process-based LCAs on average and specific productions of aggregates in Quebec, Canada, to understand the environmental impacts of aggregate production and consumption, their variability, and the factors of this variability. Environmentally-extended input-output (EEIO) analysis, and hybrid LCA – complementing process-based LCA with EEIO –, are not suitable approaches, as the gravel production sector is aggregated with other mineral-related activities. Thus, they do not allow to estimate the specific environmental impacts of aggregates, and even less the ones from a particular production type (Agez, 2021; de Bortoli and Agez, 2023; Stadler et al., 2018; Yang et al., 2017). Thus, following ISO 14040 and 14044 (International Organization for Standardization, 2006a, 2006b), I will develop average production models for the years 2020 and 2021 to assess the interannual variability of impacts, as well as develop models for specific sites to assess the inter-site variability. Among the five specific production units appraised, three fixed unit productions will be studied: the quarries of Laval (limestone rock), Saint Bruno (volcanic rock), and Saint Philippe (limestone). Two mobile production units will also be studied, each used in two different sites over the period considered: a mobile unit called "SDE" installed in Val-des-Monts and Shawinigan sites, and a mobile unit called "SBE" unit installed in Roxton Pond and Sainte Justine. Then, scenario analyses will be conducted to give a large overview of the potential environmental impacts of aggregates: from different types of rocks, with different electricity mixes from Europe, China, or different North American regions, from fixed vs mobile installations, or with transportation to the consumer by 12-wheel truck or trailer truck on average regional distances.

The main functional unit (FU) will be "to provide one metric ton of aggregates at the quarry gate", but I will also look at the impact of aggregate transportation to customers, as the literature review showed its importance, with the following FU: "to provide one metric ton of aggregates at customer". The system boundaries considered are from cradle-to-quarry's gate, or cradle-to-customer, and thus include all activities from blasting to crushing (+ transport to customer for the second FU). The illustration of these two system boundaries can be found in Figure 3. I detail the operations of aggregate production depending on the type of installation. After obtaining the authorization to operate, a site is prepared for operation by stripping and discovery work (out scope). Then, for a massive rock site, the next step consists of drilling and blasting. The third step consists of transporting the uncrushed stones from the pit to the crushing unit, which is either electrified or using diesel if it is fixed, and runs on diesel if it is mobile. Transport can be by trucks or electric conveyor belts. In the case of a mobile production, the crushing unit is located near the working face: the main transport on-site takes place after the crushing, by truck. The greater the production of a site, the more the transport distance between the working face and the crushing site or the quarry's gates will increase over time. It should also be noted that a quarry is often a multifunctional site, where virgin aggregate ("stone", "pebble") and recycled aggregates will be produced, and aggregates and material from earthwork will be received to be stored (storage platform). At the end of the quarry operation, the redevelopment of the land will not be modeled, considering it is not a common practice in Quebec. Finally, the flows considered in the inventory within the system boundaries will be as comprehensive as allowed by the accessible data.

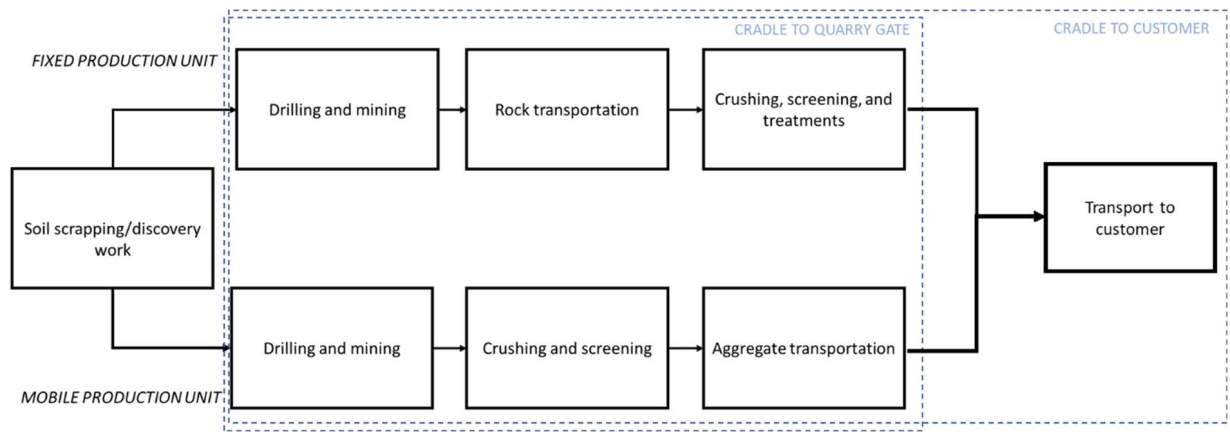

**Figure 3 Life cycle stages of aggregate production - fixed versus mobile production unit – and system boundaries of the two FU**

Alluvial installations, also called "sandpits" in Quebec, are different from hard rock production sites: these sites are often in the water and consist of deposits of the glacial eras from where alluvial rocks, also called round rocks or loose rocks, are extracted. In Quebec, they are found in Gaspésie and Bromont. Outside of Quebec, deposits may be underwater. For this type of production, operations also start by stripping the site, then exploiting the deposit with a cable or bucket shovel, or with a vacuum system. On fixed installations, a conveyor belt of variable size (from 1 to 6 km) then transports the material before processing. Extraction of loose rocks for aggregate production is increasingly rare (about 2% of Eurovia's production in 2020 and 2021 according to the data processed as part of the project) and thus not modeled specifically. But the specificity of their production will be considered in Eurovia's average production models, via their specific energy consumption and the absence of explosive usage.

## 2.2. Method to develop inventories

### 2.2.1. Background database

As a background database, I will use Ecoinvent v3.7 to develop my models, with the most common cut-off by classification allocation. The allocation type gives rules to allocate the impacts of multi-functional systems and recycling burdens and benefits. The type of allocation has been showed having potential large consequences on the results of an LCA, especially when recycling is important on the supply chain (Allacker et al., 2017). This study being attributional, I thus dismissed the consequential allocation approach. The alternative attributional allocation at the point of substitution (APOS) would probably not affect too much the results as recycling is not strongly involved along the supply chain of virgin aggregate production.

*2.2.2. Data collection and completeness*

The data are collected within Eurovia's, in production records, analytics, and interviews with quarry managers and production directors from the Eurovia Quebec subsidiary conducted in 2021-2022. It covers the entire production of aggregates by Eurovia in Quebec, from around 30 sites, with a majority of fixed production units, and some mobile installations including five Eurovia's mobile units and complementary rented mobile units. The sites produced 6.9 million metric tons (Mt) of virgin aggregates in 2020 and 9.4 Mt in 2021 (resp. 8.7 and 12.3 Mt of all kinds of aggregates). The tonnage of aggregates of all kinds produced in Quebec annually is estimated at 107 Mt, including virgin and recycled aggregates (Association des constructeurs de routes et grands travaux du Québec, n.d.). Extrapolating from Quebec's annual production and the production mix of Eurovia – including around 77% of virgin aggregates and 23% of recycled aggregates –, I estimate that Quebec produces 82 Mt of virgin aggregates and 25 Mt of recycled aggregates, and that the company thus supplies between 8 and 12% of Quebec's virgin aggregate.

*2.2.2. Data collection prioritization*

To prioritize data collection in order to enhance LCI quality and LCA results' robustness, I screen major contributors of the ecoinvent regionalized process for Quebec aggregates -

"*gravel, crushed, CA-QC, production*", calculating the impacts using the TRACI 2.0 characterization method (Bare, 2011). TRACI stands for "Tool for Reduction and Assessment of Chemicals and Other Environmental Impacts". This hotspot analysis is performed on Simapro. The off-road diesel (ORD) consumed by the machinery is a major contributor (contributing to 70% of the depletion of the ozone layer generated by the aggregate production activity, 37% to the impact on climate change, 78% to smog, 48% to acidification, 15% to eutrophication, 42% of the respiratory effects, 63% of the depletion of fossil fuels), just like the amortization of premises with metal frames (21% on the depletion of the ozone layer, 30% of the impact on climate change, 17% from smog, 32% from acidification, 29% from eutrophication, 26 and 32% from carcinogenic and non-carcinogenic health impacts, 28% from respiratory effects, 21% from ecotoxicity, 26 % of fossil fuel depletion). Heat consumption presents notable but often non-major contributions (4% of the depletion of the ozone layer, 16% of the impact on climate change, 10% of acidification, 16% of eutrophication, 27 and 7% of the carcinogenic and non-carcinogenic health impacts), as the amortization of the machinery (4% of the depletion of the ozone layer, 6% of the impact of climate change, 3 % smog, 6% acidification, 19% eutrophication, 34% carcinogenic health impact, 12% respiratory effects, 34% ecotoxicity). The amortization of the conveyor is also reflected in a few indicators (6% eutrophication, 15 and 7% carcinogenic health impact, and 6% ecotoxicity). Mine infrastructure, which includes the impacts of initial site preparation for rock extraction, contributes significantly to eutrophication (11%) carcinogenic and non-carcinogenic health impacts (8 and 16%), and ecotoxicity (24%). Electricity consumption has a secondary impact on most indicators, except on ecotoxicity (11%). In conclusion, the new aggregate LCA models developed in this study will have to focus on the validation of the energy consumption data linked to the production of aggregates and the improvement of the amortization of the quarries' modeling, in particular that of the premises and the machinery. I will also dig into the blasting stage, which has rarely been dealt with so far in published studies, and whose consumption is variable in the studies considering it, as showed in the literature review.

### *2.3. Life cycle inventories*

#### *2.3.1. Blasting*

*Explosive consumption by type of rock*

According to data from Quebec sites, explosive agent consumption depends on the type of rock: these consumptions per cubic meter of exploded rock have been collected based on Eurovia's activities and are shown in Table 4 (see SM for more details).

**Table 4 Consumption of explosives to blast different kinds of rocks**

| Type of rock | Explosive consumption | |
|---|---|---|
| | kg/m³ of rock | kg/t of rock |
| Limestone rocks | 0.45 | 0.17 |
| Dolomitic rocks and slate | 0.70 | 0.26 |
| Hard rocks (volcanic, sandstone) | 1.00 | 0.37 |

*Explosive consumption on average and by installation type*

I create representative explosive consumption models per type of Eurovia's Quebec production calculated on the product technology ratios and the associated rock types (see SM). Resulting explosive consumptions for average Quebec productions in 2020 and 2021 are showed in Table 5.

**Table 5 Explosive consumption depending on the type of installation**

| Type of production | Explosive consumption (kg/kg of rock) | |
|---|---|---|
| | *2020* | *2021* |
| AVERAGE QUEBEC PRODUCTION | 2.77E-04 | 2.76E-04 |
| MOBILE PRODUCTION | 2.97E-04 | 2.93E-04 |
| FIXED PRODUCTION | 2.67E-04 | 2.75E-04 |

*Site-specific consumptions*

Consumption of explosives for the specific sites of St Bruno (metamorphic rock, similar to volcanic rock), Laval (sedimentary rock, i.e. crystalline limestone), and Saint Philippe (same as Laval's rocks), as well as for the operations of SBE – operating at 50 % at Roxton Pond (very abrasive sandstone) and 50% at Sainte Justine (dolomite) - and SDE mobile units – operating at 50% at Val-des-Monts and 50% at Shawinigan (granitic rock, i.e. volcanic) - for 2020 were also calculated (Table 6). Inter-site explosive consumption variability is shown to be high compared to inter-technology variability (i.e., fixed vs. mobile crushing).

**Table 6 Site-specific explosive consumptions and type of rock**

| *Site* | *Type of rock* | *Explosive consumption (kg/kg)* |
|---|---|---|
| Laval | Limestone | 1.67E-04 |
| Saint Bruno | Volcanic rock | 3.70E-04 |
| Saint Philippe | Limestone | 1.67E-04 |
| SDE - Val des Monts/Shawinigan | Volcanic rock | 3.70E-04 |
| SBE - Roxton Pond/Sainte Justine | 50% limestone- 50% dolomite | 2.13E-04 |

*2.3.2 Energy and water flows*

*Foreground data: data records and calculations*

Production data were collected by production zone of Eurovia Quebec subsidiary. First, Eurovia's environmental reporting for the period from 10/2019 to 9/2020 (called 2020) was obtained from the environmental department of Eurovia group. These consumptions include standard diesel, gasoline, light fuel oil (LFO), natural gas, electricity, water, and ORD. Six production zones are studied, corresponding to the group's materials production "regions": Gaspesie quarries, North Materials, Aggregates Sainte Clotilde, Eastern Townships quarries, RSMM quarries, and Outaouais quarries. By dividing total consumptions by the tonnage produced, I calculate the consumption per ton (Table 7). Details on data and calculations can be found in the supplementary material (SM) provided with this article.

**Table 7 Consumption flows per functional unit and production type**

| Region | Diesel (L/t) | Gasoline (L/t) | LFO (L/t) | Natural gas (m3/t) | Electricity (kWh/t) | ORD (L/t) | Water (m3/t) |
|---|---|---|---|---|---|---|---|
| Fixed | 1.87E+1 | 4.94E-1 | 6.36E-4 | 4.25E-3 | 1.95E+0 | 4.52E-3 | 1.31E-5 |
| Mobile | 8.43E+0 | 0 | 0 | 0 | 9.40E-1 | 4.61E0 | 1.31E-5 |

Globally, diesel is used for stationary heavy machinery, such as fixed crushing unit generators, while gasoline is used for staff road trips (foreman vans and others). LFO is scarcely consumed for the heating of office buildings and workshops. Electricity is used by the pumps and in crushing activities in many fixed production sites such as in the quarries of Laval, St Bruno, St Philippe, and Bromont. Finally, ORD is used for mobile machinery: trucks, loaders, and mobile crushing units.

The water consumed comes mainly from the environment: the consumption of running water is rare as sites are generally not connected to the water network. It will be considered that the water consumed comes 75% from rainwater and 25% from groundwater based on Eurovia's experts. The water is used to reduce dust emissions: it then goes into settling ponds, and only overflows can be discharged into the environment. But water discharges are very low and meet the thresholds settled by the Ministry of the Environment. Moreover, no flocculation product is used in Eurovia Quebec's quarries.

*Background data: LCIs of fuels consumed*

Different kinds of fuels are used for different purposes. The impacts generated by the use of these fuels are due to the gas released by the combustion itself but also to the supply chain providing these fuels. The combustion impacts do not depend on the context of the combustion, but the supply chain impacts depend on the type of petroleum extracted, the technology and practices to extract, transport, and refine it. Moreover, these impacts can vary widely (Masnadi et al., 2018; Meili et al., 2018). To account for some regional aspects of these consumptions, I use several regional data for Quebec and Canada. Only GHG emissions are regionalized, following the method and using the data detailed in the supplementary material.

### 2.3.3. Dust emission

Dust emissions are not measured on-site in Quebec. Based on default emission factors recommended by the Government of Canada (US EPA, 2004), I recalculated the emission values according to the ecoinvent classification of particulate matter (PM) flows and reported them in Table 8 (see more details in SM). In Quebec, the crushing units are generally not equipped with a bag filter according to experts. The sensitivity of the environmental results to the filter will be shown in the case of the Laval quarry, whose crusher is equipped with a filter.

**Table 8 Dust emissions depending on the equipment of the crushing unit**

| *Crushing unit type* | *Dust emissions (kg/t of rock)* | | |
|---|---|---|---|
| | **PM >10µm** | **PM 2.5-10µm** | **PM <2.5µm** |
| Without filter | 0.0027 | 0.0012 | 0.0006 |
| With filter | 0.0006 | 0.0007 | 0.00005 |

*2.3.4. Facilities and machinery*

*Overview*

The main piece of equipment is a crushing unit, the operation of which is detailed in Figure 4 in the case of three crushing-stage units. Three crushing stages are required for aggregates used in concrete and asphalt mixture due to the generally small size of their aggregates.

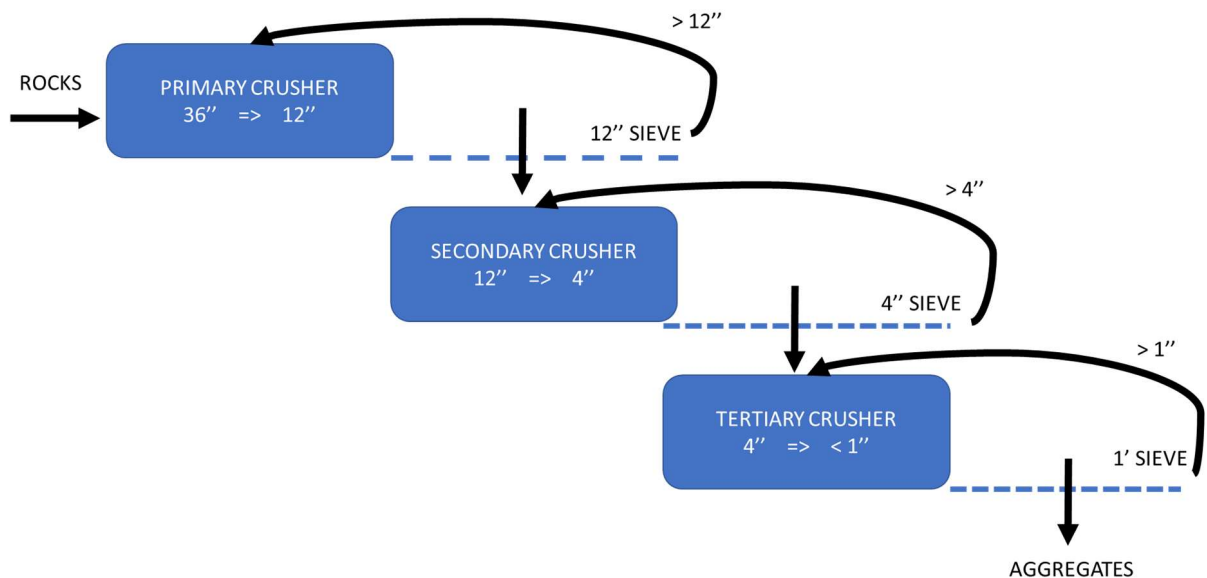

**Figure 4 Crushing unit to produce aggregates used in pavements**

This equipment can be inserted into a metal structure in certain fixed units and is mounted on tires in the case of mobile units. These structures are not metal buildings as modeled in ecoinvent gravel production process, and the metal masses of Eurovia facilities will be instead accounted for within the machinery amortization. There are no concrete premises on the quarries contrary to what is modeled in ecoinvent, but the fixed crushing units have concrete slabs to accommodate the elements of production equipment. At the EoL, the metals from the wear parts are 100% recycled. The models developed for Quebec's fixed and mobile crushing units are detailed below.

*Mobile units*

The total weight of the different elements of a mobile crushing unit is 154.5 metric tons of steel, based on the SBE and SDE units. The description of its composition is detailed in the supplementary material. Apart from the crushers and sieves, a container is linked to the generator to operate the various crushing and sieving equipment. A second vehicle (45-foot semi-trailer) serves as a temporary office for employees, a warehouse to store replacement parts, and a repair shop for equipment. The lifespan of the various materials can be estimated at around 25 years. The consumption of steel parts linked to the amortization of the crushing unit finally calculated is equal to 14.4 g of industrial machinery amortized in capital per ton of aggregate produced, for the production of 430000 metric tons per year. As done in ecoinvent's aggregate processes and due to lack of data from Eurovia's suppliers, I chose the ecoinvent process "*Industrial machine, heavy, unspecified {RoW}| market*" to model this amortization.

Wearing parts consumed by the two mobile units – SDE and SBE - were compiled (Table 9). The data was provided by the operation site managers. The differences in maintenance between the two units are explained by the nature of the rock: the softer the rock, the lower the maintenance. Conversely, a more abrasive rock wears out the machinery more quickly. Parts are made of manganese steel, i.e. 18% of manganese and the rest of steel. Manganese hardens under impacts, while steel is not strong enough for these units. Manganese steel, also called Hadfield steel or mangalloy, does not exist in ecoinvent. Therefore, I will only consider the material impact of manganese and steel manufacturing, without considering the impact of mixing the two materials. I calculated 36.6 and 49.3 mg of wear parts consumption per kilogram of crushed stone respectively for the SDE and SBE plants, i.e. an average of 42.9 mg/kg of aggregate produced. Metal recycling is not accounted for, as mobile units are mainly owned by subcontractors, meaning that Eurovia does not control the recycling. This is a conservative approach.

**Table 9 Annual maintenance of two Quebec mobile units – SBE and SDE units**

| Wear parts | SDE units | SBE units | Mass (kg/unit) | SDE maintenance (mg/kg) | SBE maintenance (mg/kg) |
|---|---|---|---|---|---|
| Primary sleeve | 3 | 2 | 1 500 | 10.47 | 6.98 |
| Primary hammer | 2 | 2 | 1 665 | 7.74 | 7.74 |
| Secondary mantle | 3 | 2 | 1 160 | 8.09 | 5.40 |
| Secondary dry hammer | 3 | 2 | 1 390 | 9.70 | 6.47 |
| Third mantle | 4 | 3 | 615 | 5.72 | 4.29 |
| Third hammer | 4 | 3 | 815 | 7.58 | 5.69 |

*Fixed units*

For accessibility reasons, data collection was carried out on three representative fixed plants: Laval, Saint-Bruno, and Saint-Philippe. The characteristics of these sites were given by Denis Bouchard, and all the following elements were accounted for: basin, concrete infrastructure, industrial machinery (crusher, conveyor, etc.) (see details in supplementary material). The final inventories to be considered per ton produced are calculated according to the production and lifespan of each site. The inventories of the three quarries, in terms of LULUC, infrastructure and equipment amortization, are presented in Table 10. In Simapro, these inventories have been recalculated for each kilogram produced. Note that I modeled three sites considered representative of Quebec, although significant economies of scale make the mass of steel used per ton produced decrease rapidly on the largest facilities, and, vice versa, increase on the small ones. Moreover, while LULUC effects are not accounted for in the characterization method used in this study, it is important to include them in the inventories to allow future assessments quantifying these phenomena.

**Table 10 LULUC, infrastructure and equipment amortization for fixed crushing unit quarries**

|  | Quarries | | |
|---|---|---|---|
|  | **Laval** | **St-Bruno** | **St-Philippe** |
| Annual production (kt) | 1200 | 1100 | 775 |
| Site lifespan (years) | 40 | 30 | 65 |
| **Land transformation** (m²/t aggregates) | | | |
| Ponds | 1.15E-03 | 8.71E-05 | 7.94E-04 |
| Rest | 3.51E-01 | 1.59E-02 | 1.65E-02 |
| **Land use** (m².an/t aggregates) | | | |
| Ponds | 4.58E-02 | 2.61E-03 | 5.16E-02 |
| Rest | 1.40E+01 | 4.77E-01 | 1.07E+00 |
| **Infrastructure amortization** (m3/t aggregates) | | | |
| Cement concrete 30-32 MPa | 3.18E-05 | 2.99E-05 | 1.49E-05 |
| **Equipment amortization** | | | |
| Industrial machinery + metal frame (t/t aggregates) | 1.07E-05 | (1.07E-05) | (1.07E-05) |
| Conveyor belt (m/t aggregates) | 4.E-05 | 5.E-05 | 2.E-05 |

For the maintenance of the fixed unit, the parts worn annually as well as their number, the total mass to replace per kilogram of aggregate produced, and finally composition are indicated in Table 11. Data collection was carried out in Laval's quarry. The maintenance inventories per kilogram of aggregate produced are calculated according to the annual production of Laval's quarry and presented in Table 11. As declared by Denis Bouchard, 10% of the conveyor needs replacement per year on average. The 10% tungsten included in the liners of the secondary crusher is neglected because no process is available in ecoinvent v3.7. On the other hand, it can be taken into account in the future with ecoinvent v3.8 using one of the two new tungsten processes. The function of the different spare parts is explained in the supplementary material.

**Table 11 Fixed unit maintenance**

| *Wear parts* | *Number* | *Mass (kg/year)* | *Maintenance (mg/kg aggregate)* | *Composition* |
|---|---|---|---|---|

| | | | |
|---|---|---|---|
| Feeder | 1 | 1600 | *1.33E+00* | Steel |
| Primary sleeve | 1 | 2041 | *1.70E+00* | Steel, 18% manganese |
| Primary hammer | 3 | 3402 | *2.83E+00* | Steel, 18% manganese with 2% chrome |
| General liners | 1 | 1588 | *1.32E+00* | Steel |
| Secondary shield | 12 | 1905 | *1.59E+00* | 20% ceramic, 80% manganese |
| Secondary dry hammer | 8 | 6169 | *5.14E+00* | Steel, 18% manganese with 2% chrome |
| Liners for secondary crusher | 24 | 1043 | *8.69E-01* | 90% steel, 10% tungtene carbonate |
| External screws' pads | 150 | 680 | *5.67E-01* | Steel |
| Internal screws' pads | 25 | 283 | *2.36E-01* | Steel |
| **Spare part** | Ratio | Maintenance (m/an) | | Composition |
| Conveyor belt replacement | 10% | 1.40E-07 | | Mixed (see ecoinvent process) |

*2.3.5. Aggregate downstream transportation*

Aggregate transport includes the entire life cycle of the transportation service, i.e., manufacturing and maintenance of the truck as well as fuel production and combustion. The vehicle EoL is excluded as the impact is negligible (de Bortoli et al., 2017). Because the generic truck transportation processes in ecoinvent are not representative of the construction practices, I developed new tailored models.

*Average loads and distances*

Upstream transport - between the pit and the processing facility - is considered in the site energy consumption data presented above. Data on the downstream transport of aggregates - average supply distances from quarries to customers, type of truck used, and average full load - were gathered. The information is summarized in Table 12. In addition, no double freight can generally be set up. Thus, the new aggregates transport models will consider trucks empty returns. The average transport distance for aggregates is 16.9 km between quarries and Eurovia's customers in Quebec. Average full loads vary by the quarry, with a 2020 production weighted average of 32.8t, but an average of resp. 35.5 and 22.0 tons for semi-trailers and 12-wheel trucks.

**Table 12 Transportation characteristics for aggregates, from the quarry to the consumer, by region**

| Production zone | Vehicle type | Average distance from quarry to consumer (km) | Average full load (t) |
|---|---|---|---|
| GASPESIE QUARRIES | Semi-trailer | 45.5 | 35 |
| NORTH MATERIALS | 12-wheel trucks & semi-trailer | 3.1 | 22 |
| AGGREGATES SAINTE-CLOTILDE | Semi-trailer | 32.9 | 36 |
| EASTERN TOWNSHIP QUARRIES | Semi-trailer | 6.7 | 34 |
| RSMM QUARRIES | Semi-trailer | 19.9 | 37 |
| OUTAOUAIS QUARRIES | Semi-trailer | 7.9 | 34 |
| **QUEBEC WEIGHTED AVERAGE** | | **16.9** | **35.5** |

*Fuel consumption*

The fuel consumption for the trucks used by Eurovia in Quebec have been collected, and regulated emissions (CO, $CO_2$, NOx, HC, PM) of heavy trucks have been recalculated based on real driving cycles, current technologies, and the consumption of the specific trucks used to transport aggregates (see SM). Final fuel consumption and emissions of trucks to transport aggregates in different operating conditions in Quebec are presented in Table 13.

**Table 13 Consumptions and emissions of aggregate trucks in different operating conditions**

| | | Consumption (L/100km) | CO | CO2 | HC | NOx | PM |
|---|---|---|---|---|---|---|---|
| Semi-trailer truck | Full, average | 50 | 1.19E+00 | 1.34E+03 | 1.20E-01 | 2.83E+00 | 4.82E-02 |
| | Empty, average | 39 | 9.30E-01 | 1.04E+03 | 9.40E-02 | 2.21E+00 | 3.76E-02 |
| | Double trip (empty return) | 89 | 2.12E+0 | 2.38E+3 | 2.10E-1 | 5.04E+0 | 8.60E-2 |

| 12-wheel truck | Rural, average | 40 | 9.54E-01 | 1.07E+03 | 9.64E-02 | 2.27E+00 | 3.86E-02 |
| | Urban, average | 52 | 1.24E+00 | 1.39E+03 | 1.25E-01 | 2.95E+00 | 5.01E-02 |
| | Double trip (empty return) | 92 | 2.19E+0 | 2.46E+3 | 0.22E-1 | 5.21E+0 | 8.90E-2 |

*2.4. LCIA*

As recommended in the ISO 21930 standard relating to LCAs of construction products in North America (International Organization for Standardization, 2017), I use the TRACI characterization method (v2 1.05) to calculate potential impacts to the environment. The calculations are carried out using the LCA software Simapro 9.1.0.11.

# 3 Results and interpretation

*3.1. Hotspot analyses*

This section aims at highlighting the main contributors to the environmental impacts of aggregate production.

*Fixed production*

Figure 5 presents the hotspot analysis of the average 2021 fixed aggregate production from cradle-to-gate. It highlights the major contribution of blasting (in red) on the final impact of the aggregate at quarry's gates: from 18% of the total impact on ozone depletion to 86% on acidification, and 42% of the carbon footprint. The regular diesel, globally burned by generators (in black) is the second biggest contributor overall – accounting for a substantial part of the impacts, especially on global warming (47% of the impact), ozone depletion (75%), and fossil fuel depletion (70%). On health impacts (carcinogenics and non-carcinogenics) as well as ecotoxicity on the other hand, the conveyor belt manufacturing and maintenance are the second most important contributor. Its maintenance brings more impact than its manufacturing. Recycling of machinery spare parts allows for reducing the environmental impacts on health (both for carcinogenic and non-carcinogenic diseases, resp. by 8 and 4%), and ecotoxicity (-8%). The production of manganese (in purple) for the different spare parts of the crushing unit generates one-quarter of the carcinogenic impacts, especially due to the maintenance of the hammers. Dust generated on-site only contributes to respiratory effects (15%): scopes 2 and 3 emissions are far more important than scope 1 emissions, according to the GHG protocol definition of scopes (WBCSD and WRI, 2011). Interestingly, the original machinery amortization (in blue), which is the only equipment considered by the original ecoinvent models (Kellenberger et al., 2007), only brings limited impacts (0 to 5% maximum, on ecotoxicity). Just as electricity, accounting for 0 to 6% of the impact, and ORD, mostly consumed by trucks on site.

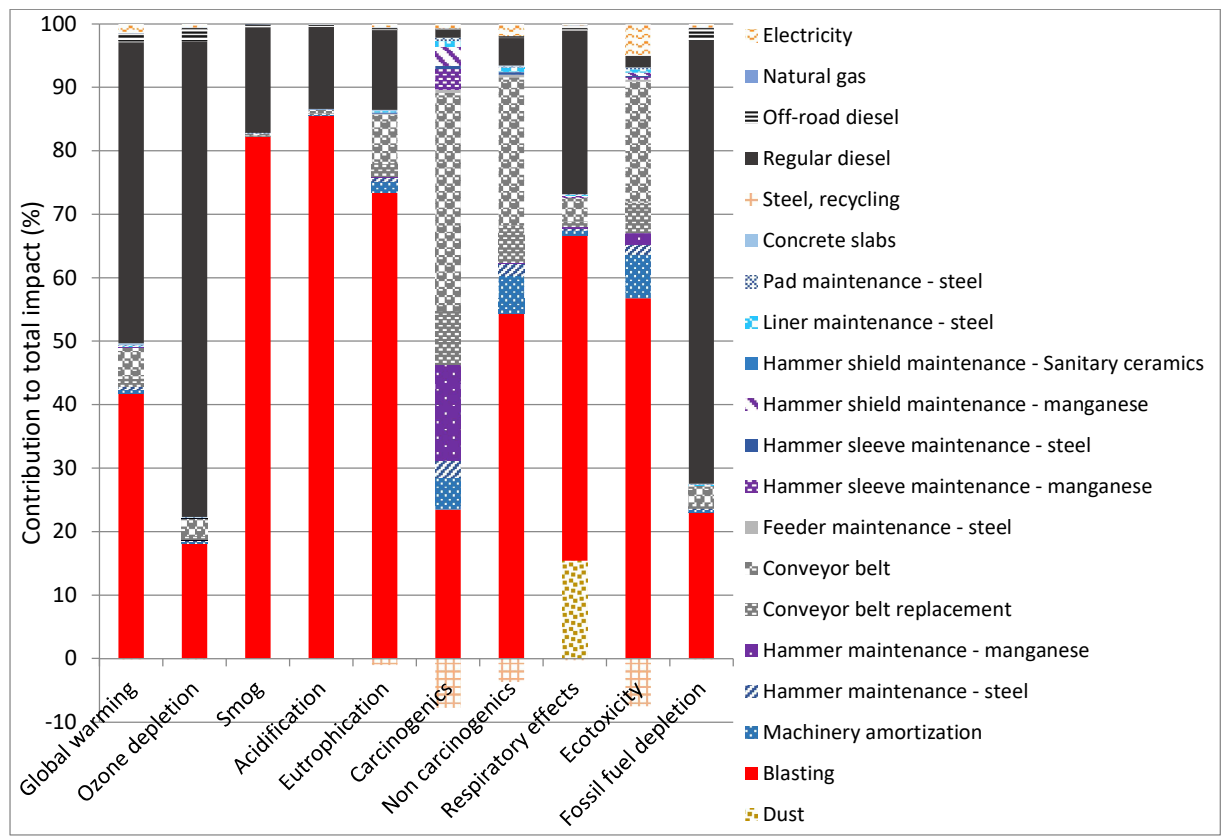

**Figure 5 Hotspot analysis of aggregate production activities for Eurovia Quebec in 2021, for fixed production**

*Mobile production*

Figure 6 presents the hotspot analysis of the average 2021 mobile aggregate production of Eurovia in Quebec. The blasting (in red), with 25 to 90% of the impact contribution, is again the most important contributor on most indicators. Then, the regular diesel, generally burned in the crusher generator, is the second biggest contributor (13-46%) on all the indicators, to the exclusion of health impacts and ecotoxicity. The ORD, either used by the mobile crushers or on-site trucks, then brings the most important impacts (0-25%), while the contribution was very low in the case of fixed production. The carcinogenics impact category displays very specific contributions: 65% of the impact is explained by the machinery amortization and maintenance, mainly by the manganese production for the maintenance of the hammer and the sleeve (51%, in purple), and the rest by the manufacturing of steel (in blue). While the crusher's spare parts are only made of 18% of manganese and 82% of steel, manganese brings the biggest contribution due to its high impact. Steel still brings a substantial impact on non-carcinogenic health impacts and ecotoxicity, with resp. 21 and 19% of the total impact.

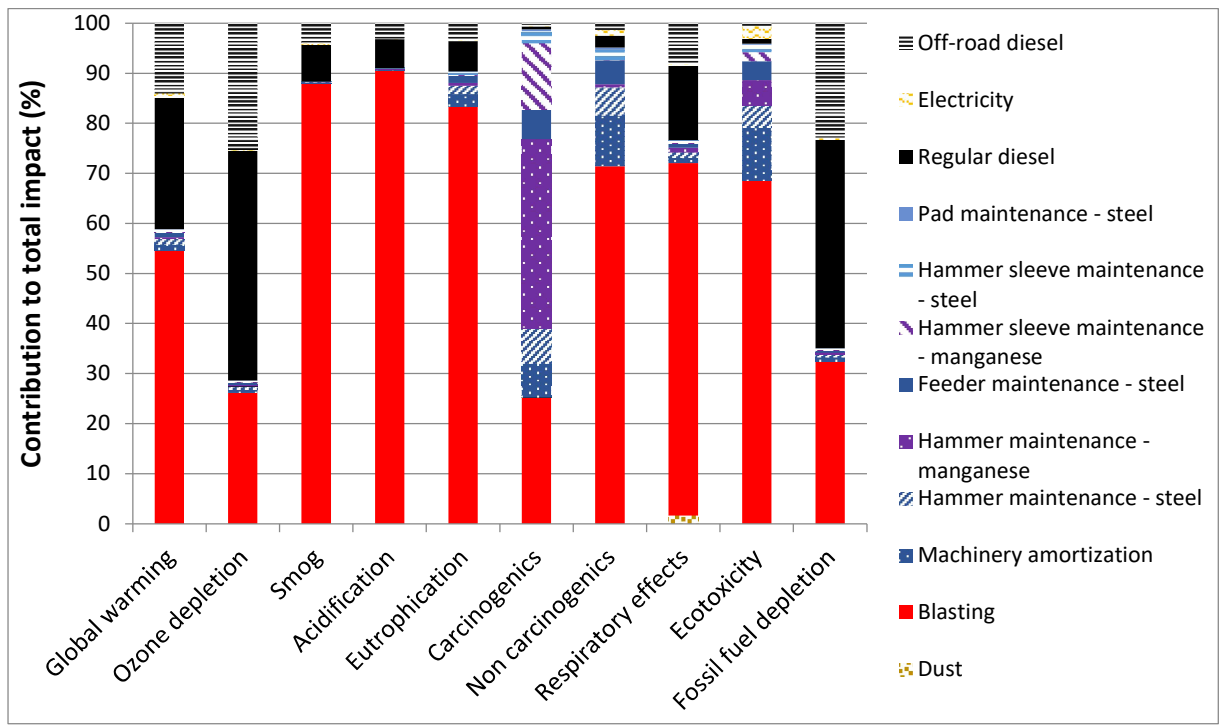

**Figure 6** Hotspot analysis of aggregate production activities of Eurovia Quebec in 2021, for mobile production

## 3.2. Quebec's aggregates impacts

The environmental impacts of different aggregates produced in Quebec by Eurovia are presented in Table 14 for each TRACI's impact category. Their comparison is shown in Figure 7. Eurovia's aggregates produced in Quebec emitted on average 2.97 kgCO$_2$eq/t in 2020 and 3.03 kgCO$_2$eq/t in 2021. This slight increase can be explained by slightly harder rocks mined in 2021 compared to 2020, and to the larger part of fixed-produced aggregates in 2021, these aggregates being more impacting than mobile-produced aggregates (see below).

Site-specific LCAs show aggregate carbon footprints ranging between 2.28 and 3.59 kgCO2eq/t. The aggregates produced with the SBE unit have the lowest impact overall, mainly due to the low consumption of explosives in 2020 related to the nature of the rock exploited. St Bruno quarry's aggregates show the highest impact on many indicators including climate change, due to its high explosive consumption and high consumption of gasoline and ORD.

A factor of two exists between the most and the least impacting aggregates on most of the impact categories. Let's note that the use of filters on crushing units (Laval's crusher and mobile units) does not mainly explain the respiratory effects, as aggregates from the SDE mobile unit are slightly more impacting that the average aggregates in 2020 and 2021. Inter-site variation is quite important in terms of environmental impacts of aggregate production on a cradle-to-gate perimeter, even more than what has been shown in previous studies such as the one by Jullien et al. on three different French quarries (2012).

**Table 14 Environmental impacts of different aggregates in Quebec by impact category**

| Impact category | Unit | Average. 2021 | Average. 2020 | St Bruno | St Philippe | Laval | SBE | SDE |
|---|---|---|---|---|---|---|---|---|
| Global warming | $kgCO_2eq$ | 3.03E+00 | 2.97E+00 | 3.59E+00 | 2.51E+00 | 2.86E+00 | 2.28E+00 | 3.07E+00 |
| Ozone depletion | kgCFC-11eq | 4.24E-07 | 4.13E-07 | 4.58E-07 | 3.89E-07 | 4.60E-07 | 3.26E-07 | 3.76E-07 |
| Smog | $kgO_3eq$ | 2.83E+00 | 2.82E+00 | 3.58E+00 | 1.85E+00 | 1.94E+00 | 2.15E+00 | 3.48E+00 |
| Acidification | $kgSO_2eq$ | 1.18E-01 | 1.18E-01 | 1.51E-01 | 7.60E-02 | 7.90E-02 | 8.95E-02 | 1.47E-01 |
| Eutrophication | kgNeq | 1.11E-02 | 1.10E-02 | 1.43E-02 | 7.75E-03 | 8.16E-03 | 8.31E-03 | 1.32E-02 |
| Carcinogenics | CTUh | 4.78E-07 | 4.80E-07 | 5.53E-07 | 3.94E-07 | 4.20E-07 | 4.64E-07 | 6.19E-07 |
| Non carcinogenics | CTUh | 6.91E-07 | 6.79E-07 | 9.41E-07 | 5.58E-07 | 5.93E-07 | 5.02E-07 | 7.48E-07 |
| Respiratory effects | kgPM2.5eq | 7.14E-03 | 6.97E-03 | 9.06E-03 | 5.95E-03 | 2.09E-04 | 4.95E-03 | 7.30E-03 |
| Ecotoxicity | CTUe | 5.45E+01 | 5.40E+01 | 7.21E+01 | 4.21E+01 | 4.54E+01 | 4.31E+01 | 6.41E+01 |
| Fossil fuel depletion | MJsurplus | 4.06E+00 | 3.96E+00 | 4.47E+00 | 3.64E+00 | 4.27E+00 | 3.11E+00 | 3.71E+00 |

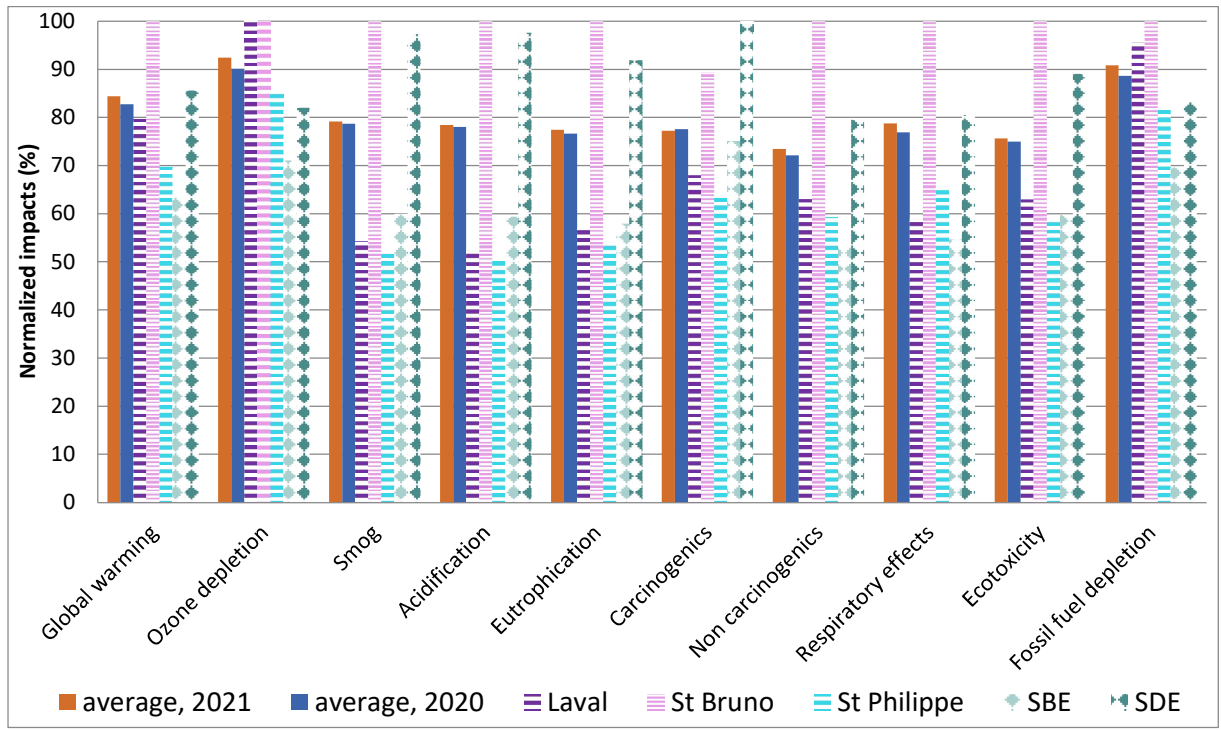

**Figure 7** Comparison of the environmental impacts of Quebec's average and site-specific aggregate productions

### 3.3. Scenario analyses

*3.3.1. Nature of rocks*

Figure 8 presents the sensitivity of the environmental impacts of aggregates to the consumption of explosives, that itself depends on the nature of the rocks exploited. These impacts have been normalized based on the most impacting aggregate. It shows a high sensitivity of the impact to the type of rocks – with a reduction of 13 to 49% of the impact between volcanic & sandstone rocks on one side (in blue), and limestone rocks on the other side (in purple), which respectively impact the most and the least the environment. The sensitivity of the results is particularly striking for air pollution-related impact categories. In particular, the nature of rock explains a difference by a factor of two in terms of smog, acidification, and eutrophication, due to the atmospheric emissions released by blasting.

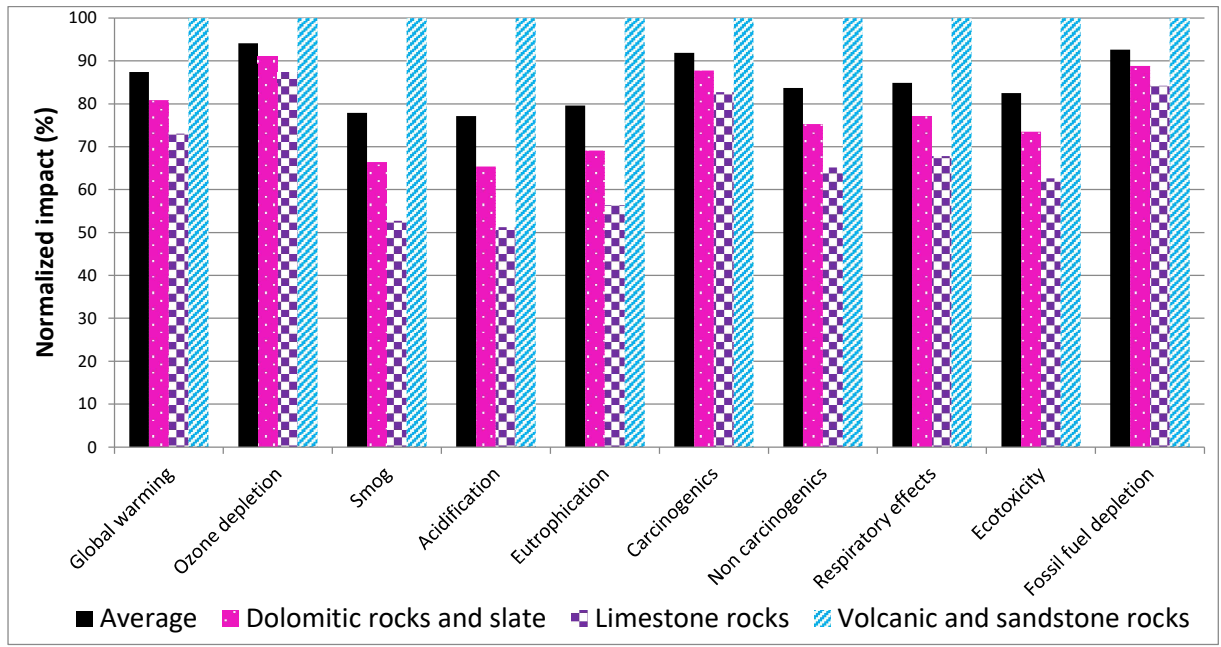

**Figure 8 Normalized comparison of 2021 aggregate fixed production impacts depending on the nature of rocks**

*3.3.2. Electricity mix*

Figure 8 shows comparison of the results of a scenario analysis where, based on the LCA model for the 2021 average Eurovia's fixed production in Quebec, the electricity mix is modified, considering the mix in other regions: other provinces of Canada – Ontario, Alberta – or with the average Canadian electricity supply, United States regions – the northeast part (NPCC electricity, for Northeast Power Coordinating Council), the western part (WECC electricity, for Western Electricity Coordinating Council) and the Midwest (MRO electricity, for Midwest Reliability Organization) -, Europe, and China. Results show an important sensitivity of the aggregate impact to the electricity mix for global warming (the least impacting aggregates emit -37% of GHG than the most impacting ones), eutrophication (-68%), non-carcinogenics emission (-50%), and respiratory effects (-45%). Results absolute values are indicated in SM, and the carbon footprints of these aggregates range between 3.28 and 4.98 kgCO$_2$eq/t, resp. in Quebec and China. Although our system boundaries are more complete than those of any study done so far – including blasting and capital goods manufacturing and maintenance -, our carbon footprint range is consistent with the literature presented in Figure 1. Canada's provinces present very variable impacts, with Alberta's aggregates displaying the worst consequences for the environment, due to a 90% fossil-based electricity mix. Within the scenario explored, Alberta's aggregates have the highest impacts on eutrophication, health effects, and ecotoxicity, while China's aggregates are the worst on global warming, smog, and acidification. On the other impact categories, each of the three zones from the US alternatively presents the highest impact. US productions rather rank in the highly impacting aggregates, except northeast production on some indicators. Quebec's aggregates are always the least impacting thanks to the very high share of hydroelectricity in its mix (94%) (Canada Energy Regulator, 2022).

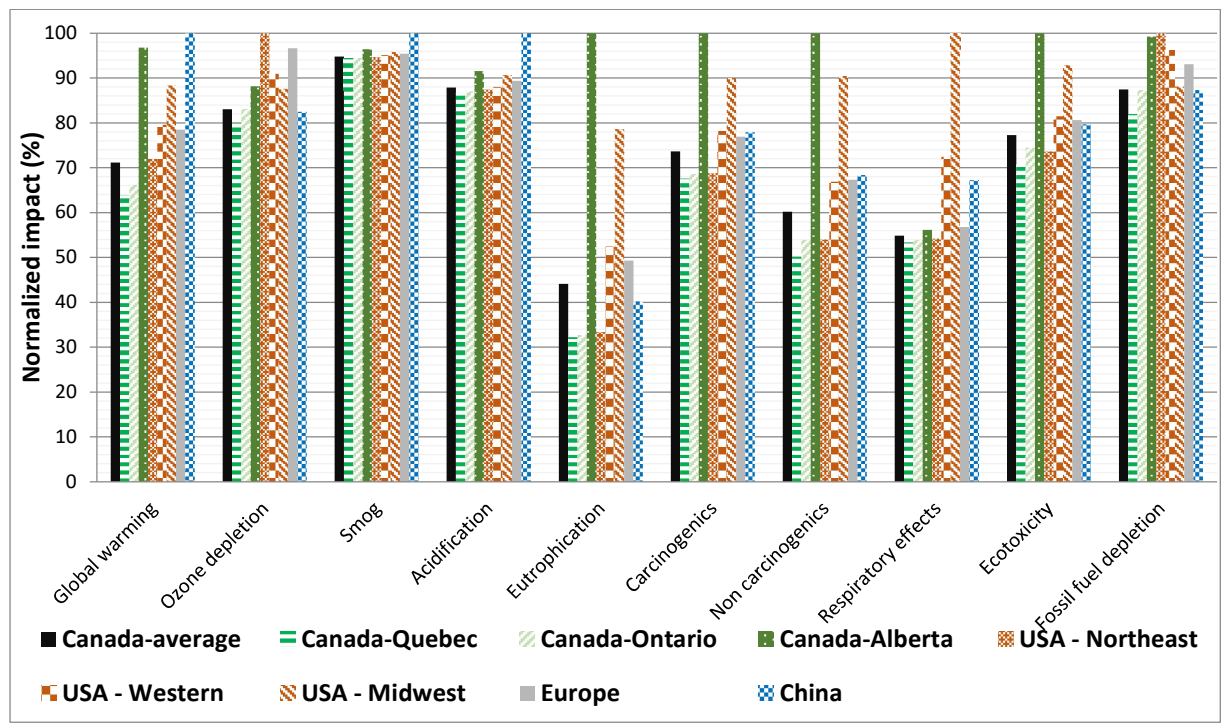

**Figure 9 Sensitivity of the aggregate's impacts to the electricity mix: comparison of Canadian provinces, US zones, European average, and China**

*3.3.3. Nature of the installation: mobile or fixed*

Figure 9 compares the impact of average, fixed, and mobile productions of aggregates by Eurovia in 2021 (2020 results are not presented, as the inter-annual variation was under 3%). Based on our models, mobile production is overall more efficient than fixed production. It especially reduces by 25% the impacts compared to the fixed production in terms of GHG emissions. Yet, the consumption of explosives has been more important for the mobile industry –2.93 $10^{-4}$ kg/kg for mobile industry and 2.75 $10^{-4}$ kg/kg for the fixed industry – due to the nature of the rocks blasted on mobile sites. But lower fossil fuel consumptions are reported on mobile sites. On the health impact categories, the higher impact of the crushing unit maintenance in the case of mobile productions compared to fixed productions is explained by the exploitation of more abrasive rocks in the case of mobile production, which is not intrinsically related to the type of installation. But as the production and maintenance of conveyor belts on fixed production sites generate significant impacts, mobile production is ultimately less impactful on these two indicators.

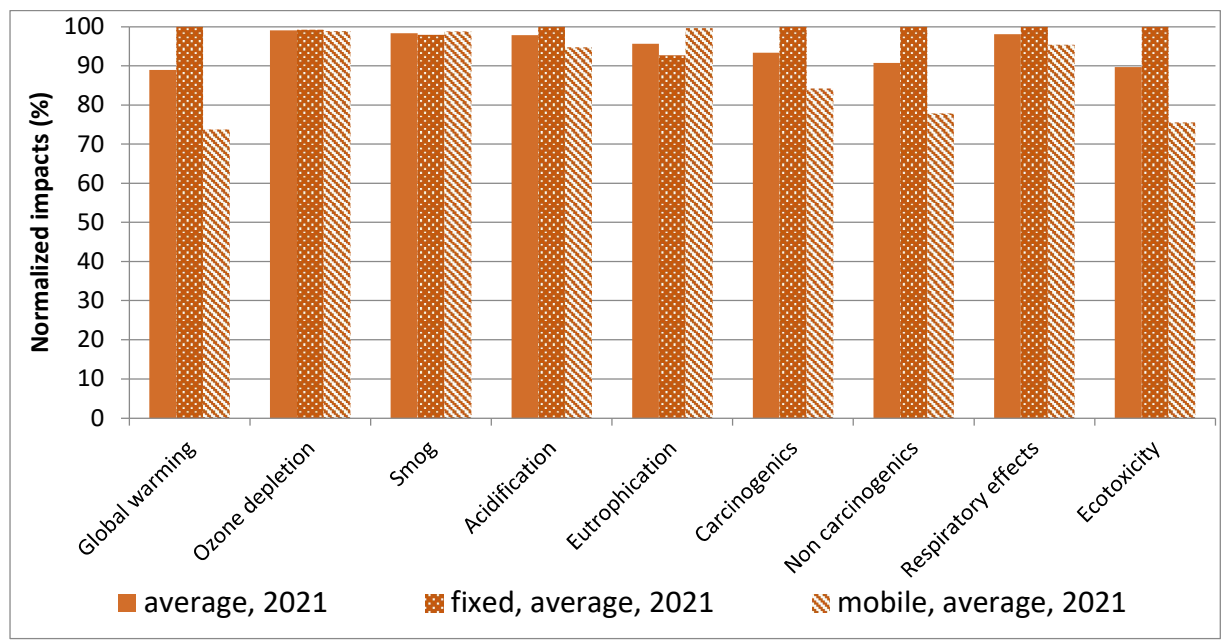

**Figure 9 Comparison of the environmental impact of average Quebec's aggregate productions with fixed versus mobile units.**

Nevertheless, the energy consumption reported by external crushing service suppliers must be questioned. Indeed, fuel consumption reported by Eurovia in its fixed sites is higher than fuel consumption reported by suppliers on mobile sites. Yet, fixed sites tend to be more electrified, normally implying a lower fuel consumption. The difference could partly be explained by shorter transportation distances from the basting site to the crusher on mobile sites, but this will need to be investigated further.

*3.3.4. Transportation to consumer*

Two kinds of trucks can be used to transport aggregates to the consumer: the most common trailer trucks, or smaller 12-wheel trucks. Moreover, distances between quarries and customers vary widely in Quebec. Figure 10 presents the variability of aggregate carbon footprints, from cradle-to-consumer, depending on the distance from the quarry's gates to the customer, the truck used, and the transportation model used. I developed LCA models for 12-wheel trucks and trailer trucks transporting aggregates in different usage conditions (urban, rural, mixed driving cycles; full load, empty, average load) in Quebec (see methodological section and SM), and compared their results to the results using the default truck transportation model from ecoinvent ("*market for transport, freight, lorry 16-32 metric ton, euro5 {RoW}*"), with doubled average distances to account for empty returns. The carbon footprint of the different options and models per ton of aggregates transported over one kilometer (tkm) is provided in the supplementary material and shows that 12-wheel trucks emit less than trailer trucks. Final results (Figure 10) show that transport to the customer can more than double the carbon footprint of the aggregate at quarry gates, in the case of the region where transportation distances are the longest (i.e., Gaspesie's region, 45.5 km in average). In the case of the shortest distance (i.e., Northern Materials area, 3.1 km in average), with 12-wheel trucks, the extra emission due to transportation only represents 14% compared to the aggregate carbon footprint at the quarry gate. On average, transportation adds 46% to the carbon footprint at the gate. Nevertheless, results show that the ecoinvent by-default model overestimates by a factor of two the emissions due to transportation. This is not due to double distances considered in this model, but to average fuel consumptions data as well as feeble average loads considered in the truck transportation models in ecoinvent: a bit less than 6 tons for the biggest truck, versus three times more for the average loaded trailer truck carrying aggregates in Quebec (35.5 tons at load, and empty returns). As ecoinvent transport processes calculate the impact per ton transported over one kilometer, the lower the load considered, the higher the impact per ton-kilometer. Tailoring the transport models in construction LCA is thus critical not to overestimate the impacts from this stage.

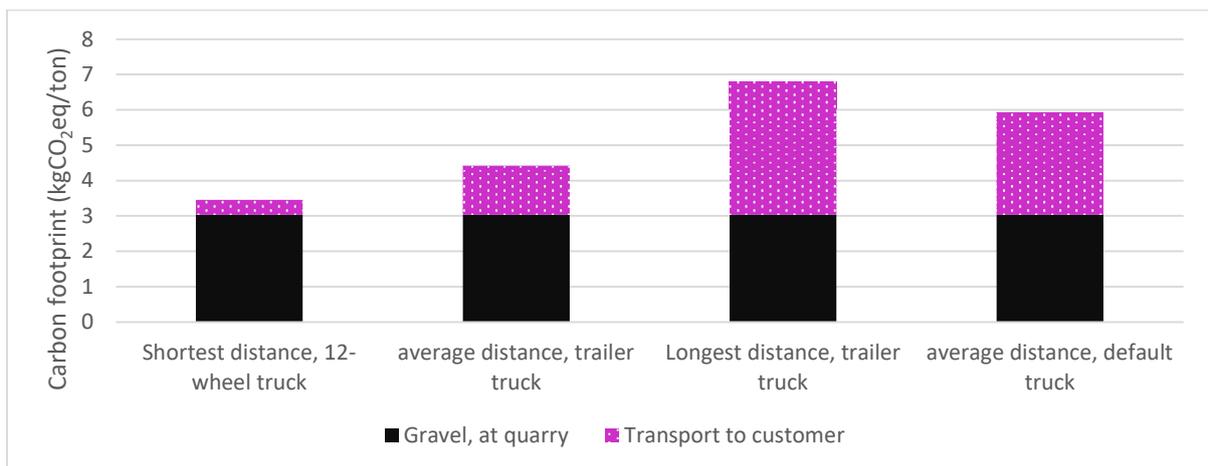

**Figure 10 Aggregate carbon footprint from cradle-to-customer, from production (in black) and transportation (in magenta)**

# 4   Discussion

## *4.1. Consistency with ecoinvent LCAs*

I will thus discuss my results through a comparison of my models with ecoinvent models based on real data collections in Switzerland, Brazil, and India (Ananth and Mundada, 2017;

Kellenberger et al., 2007; Silva et al., 2018), as these are the impacts calculable with the same TRACI characterization factors. The models in ecoinvent display various carbon footprints from cradle to gate: 2.33 and 3.4 kgCO$_2$eq/t in Switzerland (resp. from limestone and gravel, crushed, GWP100 IPCC 2021), 4.9 kgCO$_2$eq/t in Brazil, and 7.3 kgCO$_2$eq/t in India (, see Table 15 and Figure 11). The Swiss inventory displays the lowest diesel consumption of all the models compared – 14 MJ/t against 27 MJ/t for the Brazilian inventory and 30 MJ/t for the Indian inventory – while our fixed-produced aggregate inventory reports 19 MJ of diesel consumed by ton of aggregate produced. As the Swiss inventory does not account for blasting, which is the top two contributor to the carbon footprint along with diesel in my models, its carbon footprint must then be even lower than the carbon footprint range that I get in this Canadian study. But the rough machinery and building amortization model, accounting for a substantial part of the carbon footprint, explains this slightly higher carbon footprint, as well as higher consumption of LFO accounting for 8% of the Swiss carbon footprint. This quite high consumption of LFO (used in theory to heat buildings) seems unlikely based on current practices. This comparison also shows that our diesel consumption data, although in the lower range of existing LCI's diesel consumption, are consistent.

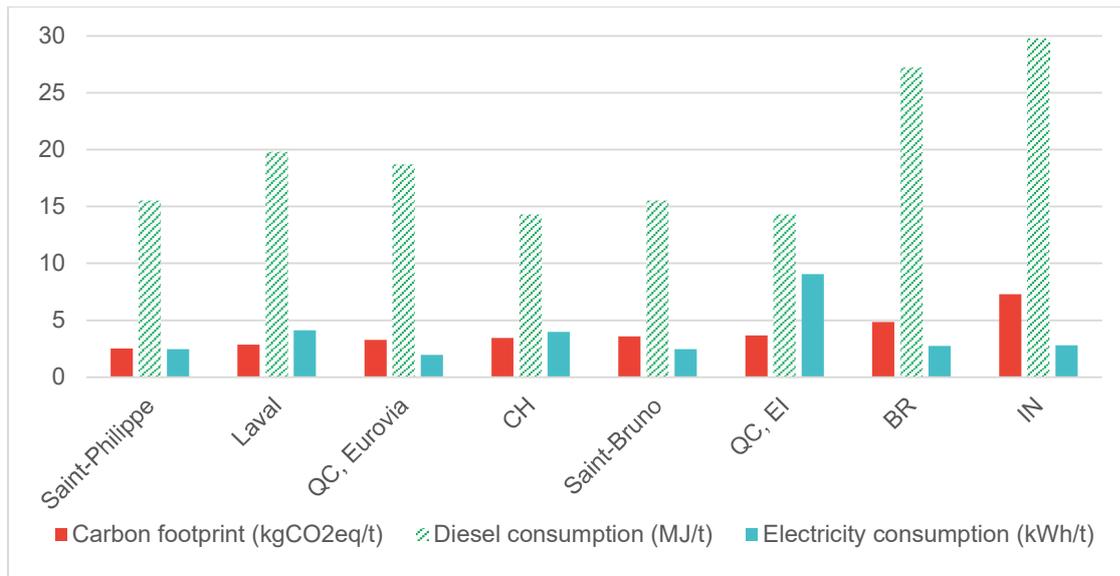

**Figure 10 Comparison of carbon footprints, diesel, and electricity consumption in various aggregate LCA models and results**

**Table 15 Comparison of carbon footprints, main consumptions, and type of rocks in various aggregate LCA models and results**

| Source | Geography | Carbon footprint (kgCO2eq/t) | Diesel consumption (MJ/t) | Electricity consumption (kWh/t) | Explosive consumption (g/t) | Type of rock | Comments |
|---|---|---|---|---|---|---|---|
| This study | Saint-Philippe | 2.51 | 15.5 | 2.47 | 167 | Limestone | |
| This study | Laval | 2.86 | 19.8 | 4.11 | 167 | Limestone | |
| This study | QC, Eurovia | 3.28 | 18.7 | 1.95 | 275 | Mixed | "gravel, fixed, average, 2021" |
| ecoinvent | CH | 3.44 | 14.3 | 3.98 | 0 | ? | "gravel, crushed" |
| This study | Saint-Bruno | 3.59 | 15.5 | 2.47 | 370 | Volcanic (hard rock) | |
| ecoinvent | QC, EI | 3.68 | 14.3 | 9.06 | 0 | ? | |
| ecoinvent | BR | 4.86 | 27.2 | 2.74 | 370 | Granite (hard rock) | |
| ecoinvent | IN | 7.28 | 29.7 | 2.79 | 70.9 | Granite and black trap stone | Basting emissions excluded |

*4.2. Nature of rocks matters the most*

Jullien et al. did not find any energy consumption difference depending on the nature of rocks: crushing requested 10.51 MJ/t (18.5 MJ/t in total) for the hard rocks they analyzed, while soft rocks requested 6.3 MJ/t on one site and 12.4 MJ/t on the other site. On average, the consumption of diesel and electricity represent 28.3 MJ of primary energy consumed per ton for the French aggregates studied in this article, against a range between 24.4 and 34.6 MJ/t in our study, and 28.6 and 46.9 MJ/t in ecoinvent. However, the results and the comparison in section 4.1 based on ecoinvent and my Canadian models show that explosive consumption - thus the nature of rocks - is one of the best explicative variables of the carbon footprint of aggregates. Hard rock aggregates tend to present higher impacts than softer rock aggregates because of higher explosive consumption, but they also wear out the machinery faster, generating higher impacts on metal-sensitive impact categories. Let's also note that the aggregate carbon footprint is secondly explained, after explosive consumption, by diesel consumption as showed in section 3.2. But it can also be substantially explained by electricity consumption in countries where electricity mixes have high carbon intensity, as the scenario analysis on electricity mix in section 3.3.2 highlights.

*4.3 Toward high-quality uncertainty assessments?*

This article illustrated the variability of aggregates impacts and explored the main contributors to different impact categories based on a comprehensive system boundary and high-quality inputs. Nevertheless, it did not quantify the other side of uncertainties, relative to stochastic errors. Indeed, uncertainties occur at each stage of an LCA (Baker and Lepech, 2009), due to both natural variability that can be missed due to narrow sampling (low completeness of the production audited) and errors due to low reliability of input data. Performing scenario analysis, as done in this study, is a recommended practice to test or generalize LCA results. More advanced mathematical methods may be more adapted to capture uncertainties, such as Monte Carlo simulations (MCS) or global sensitivity analysis methods (Saltelli and Annoni, 2010).

Yet, there is no consensus on the right tools to assess LCA uncertainties. MCS combined to the pedigree matrix uncertainty factors is the most common mathematical approach, especially used when no uncertainty can be calculated on the inputs based on real samples. Nevertheless, Heijungs stated that this method should not be used at all, and showed that this common method is lazily used as a consensus while not even the number of runs is questioned (Heijungs, 2020). This aspect will need to be investigated further to address the uncertainty of the environmental impacts of aggregates.

# 5 Conclusions

My article presents the most advanced aggregate LCA models to date, allowing a better understanding of the environmental impacts of production and transport of the most used material in the world. It shows the importance of modeling the often-forgotten blasting and machinery (crusher unit and conveyor belt), as they are major contributors to several categories of environmental impact. I also demonstrated for the first time the high sensitivity of environmental impacts to the nature of the rock exploited. First, the harder it is, the more explosive it requires, which increases the impact. Second, the more abrasive the rock, the faster it wears out crushing unit wear parts, whose maintenance especially generates tensions on health impacts as well as ecotoxicity. I showed the pronounced sensitivity of the impacts to the electricity mix, by evaluating the case of several Canadian and American areas, as well as Europe, and China : up to a factor of 3 between the least and the most impacting aggregate. Finally, I showed that truck transportation must be modeled based on construction-specific data since ecoinvent generic model overestimates the impacts of aggregate truck transportation by a factor of two. With tailored models, the average environmental burden to transport the aggregate to the consumer increases the impact of the aggregate at the quarry gates by nearly 50% in Quebec, with strong regional variability (+14 to +96%). In terms of perspectives, the development of new explosive and equipment LCIs, the monitoring of diesel consumption, machinery lifespan and wear depending on the nature of rocks, as well as on-site dust emission measures, are the next steps to improve aggregate LCI models. But in the short term, I call for completing the existing aggregate LCIs in the reference databases, especially with blasting and machinery wear, as well as customizing truck transport LCIs for construction, to assess the environmental impacts of aggregate consumption more accurately.

**Acknowledgments**: My warmest thanks go to Denis Bouchard, Vice-President of Quarries at Eurovia Quebec, for his unswerving support and dynamic supply of technical information and feedback. I also thank Ivan Drouadaine for funding and making this study possible, Marc Proteau and Amelie Griggio for hosting the project at Eurovia's Americas Technical Center, as well as Pauline Swinnen, Sylvain Delage, Thierry Roux, Julie Hébert, and Karen Bernard, for providing the primary field data and complimentary technical information to conduct these LCAs.

**Funding source and role:** this study has been funded by Eurovia's technical department to generate primary data to conduct regionalized and high-quality LCAs and develop reliable environmental transition plans.

REFERENCES

Agez, M., 2021. OpenIO-Canada. https://doi.org/10.5281/ZENODO.5553377
Allacker, K., Mathieux, F., Pennington, D., Pant, R., 2017. The search for an appropriate end-of-life formula for the purpose of the European Commission Environmental Footprint initiative. Int. J. Life Cycle Assess. 22, 1441–1458. https://doi.org/10.1007/s11367-016-1244-0
Ananth, K.P.V., Mundada, M., 2017. Life Cycle Inventories of Cement, Concrete and Related Industries - India.

Baker, J., Lepech, M., 2009. Treatment of uncertainties in Life Cycle Assessment. Stanford University.

Bare, J., 2011. TRACI 2.0: the tool for the reduction and assessment of chemical and other environmental impacts 2.0. Clean Technol. Environ. Policy 13, 687–696. https://doi.org/10.1007/s10098-010-0338-9

Bendixen, M., Iversen, L.L., Best, J., Franks, D.M., Hackney, C.R., Latrubesse, E.M., Tusting, L.S., 2021. Sand, gravel, and UN Sustainable Development Goals: Conflicts, synergies, and pathways forward. One Earth 4, 1095–1111. https://doi.org/10.1016/j.oneear.2021.07.008

Braga, A.M., Silvestre, J.D., de Brito, J., 2017. Compared environmental and economic impact from cradle to gate of concrete with natural and recycled coarse aggregates. J. Clean. Prod. 162, 529–543. https://doi.org/10.1016/j.jclepro.2017.06.057

Canada Energy Regulator, 2022. Provincial and Territorial Energy Profiles – Quebec.

Ciroth, A., 2013. Refining the pedigree matrix approach in ecoinvent: Towards empirical uncertainty factors.

Colangelo, F., Petrillo, A., Cioffi, R., Borrelli, C., Forcina, A., 2018. Life cycle assessment of recycled concretes: A case study in southern Italy. Sci. Total Environ. 615, 1506–1517. https://doi.org/10.1016/j.scitotenv.2017.09.107

de Bortoli, A., Agez, M., 2023. Environmentally-extended input-output analyses efficiently sketch large-scale environmental transition plans: Illustration by Canada's road industry. J. Clean. Prod.

de Bortoli, A., Féraille, A., Leurent, F., 2022. Towards Road Sustainability—Part II: Applied Holistic Assessment and Lessons Learned from French Highway Resurfacing Strategies. Sustainability 14, 7336. https://doi.org/10.3390/su14127336

de Bortoli, A., Féraille, A., Leurent, F., 2017. Life Cycle Assessment to support decision-making in transportation planning : a case of French Bus Rapid Transi, in: Proceedings of the Transportation Research Board 2017. Presented at the 96th Transportation Research Board Annual Meeting, Washington DC, USA.

Dias, A., Nezami, S., Silvestre, J., Kurda, R., Silva, R., Martins, I., de Brito, J., 2022. Environmental and Economic Comparison of Natural and Recycled Aggregates Using LCA. Recycling 7, 43. https://doi.org/10.3390/recycling7040043

Estanqueiro, B., Dinis Silvestre, J., de Brito, J., Duarte Pinheiro, M., 2018. Environmental life cycle assessment of coarse natural and recycled aggregates for concrete. Eur. J. Environ. Civ. Eng. 22, 429–449. https://doi.org/10.1080/19648189.2016.1197161

Fact.MR, 2023. Recycled construction aggregates market (No. Report 2890).

Faleschini, F., Zanini, M.A., Pellegrino, C., Pasinato, S., 2016. Sustainable management and supply of natural and recycled aggregates in a medium-size integrated plant. Waste Manag. 49, 146–155. https://doi.org/10.1016/j.wasman.2016.01.013

Federal Highway Administration, 2020. Production of Aggregate.

Foley, J.A., DeFries, R., Asner, G.P., Barford, C., Bonan, G., Carpenter, S.R., Chapin, F.S., Coe, M.T., Daily, G.C., Gibbs, H.K., Helkowski, J.H., Holloway, T., Howard, E.A., Kucharik, C.J., Monfreda, C., Patz, J.A., Prentice, I.C., Ramankutty, N., Snyder, P.K., 2005. Global Consequences of Land Use. Science 309, 570–574. https://doi.org/10.1126/science.1111772

Fraj, A.B., Idir, R., 2017. Concrete based on recycled aggregates – Recycling and environmental analysis: A case study of paris' region. Constr. Build. Mater. 157, 952–964. https://doi.org/10.1016/j.conbuildmat.2017.09.059

Gan, V.J.L., Cheng, J.C.P., Lo, I.M.C., 2016. Integrating life cycle assessment and multi-objective optimization for economical and environmentally sustainable supply of aggregate. J. Clean. Prod. 113, 76–85. https://doi.org/10.1016/j.jclepro.2015.11.092

Ghanbari, M., Abbasi, A.M., Ravanshadnia, M., 2018. Production of natural and recycled aggregates: the environmental impacts of energy consumption and CO2 emissions. J. Mater. Cycles Waste Manag. 20, 810–822. https://doi.org/10.1007/s10163-017-0640-2

Heijungs, R., 2020. On the number of Monte Carlo runs in comparative probabilistic LCA. Int. J. Life Cycle Assess. 25, 394–402. https://doi.org/10.1007/s11367-019-01698-4

Hossain, Md.U., Poon, C.S., Lo, I.M.C., Cheng, J.C.P., 2016. Comparative environmental evaluation of aggregate production from recycled waste materials and virgin sources by LCA. Resour. Conserv. Recycl. 109, 67–77. https://doi.org/10.1016/j.resconrec.2016.02.009

International Organization for Standardization, 2017. ISO 21930: 2017 - Sustainability in buildings and civil engineering works — Core rules for environmental product declarations of construction products and services.

International Organization for Standardization, 2006a. ISO 14040:2006 - Environmental management -- Life cycle assessment -- Principles and framework.

International Organization for Standardization, 2006b. ISO 14044:2006 - Environmental management -- Life cycle assessment -- Requirements and guidelines.

Jullien, A., Proust, C., Martaud, T., Rayssac, E., Ropert, C., 2012. Variability in the environmental impacts of aggregate production. Resour. Conserv. Recycl. 62, 1–13. https://doi.org/10.1016/j.resconrec.2012.02.002

Kellenberger, D., Althaus, H.-J., Künniger, T., Lehmann, M., Jungbluth, N., Thalmann, P., 2007. Life Cycle Inventories of Building Products (EcoInvent Report No. n°7). EcoInvent.

Korre, A., Durucan, S., 2009. Life Cycle Assessment of Aggregates (No. EVA025 – Final Report: Aggregates Industry Life Cycle Assessment Model: Modelling Tools and Case Studies).

Kulekci, G., Yilmaz, A., Ccedil;ullu, M., 2021. Experimental Investigation of Usability of Construction Waste as Aggregate. J. Min. Environ. 12. https://doi.org/10.22044/jme.2021.10309.1976

Langer, W.H., Arbogast, B.F., 2002. Environmental Impacts Of Mining Natural Aggregate, in: Fabbri, A.G., Gaál, G., McCammon, R.B. (Eds.), Deposit and Geoenvironmental Models for Resource Exploitation and Environmental Security. Springer Netherlands, Dordrecht, pp. 151–169. https://doi.org/10.1007/978-94-010-0303-2_8

Lesage, P., Samson, R., 2016. The Quebec Life Cycle Inventory Database Project: Using the ecoinvent database to generate, review, integrate, and host regional LCI data. Int. J. Life Cycle Assess. 21, 1282–1289. https://doi.org/10.1007/s11367-013-0593-1

Marinković, S., Radonjanin, V., Malešev, M., Ignjatović, I., 2010. Comparative environmental assessment of natural and recycled aggregate concrete. Waste Manag. 30, 2255–2264. https://doi.org/10.1016/j.wasman.2010.04.012

Marinković, S., Radonjanin, V., Malešev, M., Lukic, I., 2008. Life Cycle Environmental Impact Assessment of Concrete, in: Sustainability of Constructions: Integrated Approach to Life Time Structural Engineering; Proceedings of Seminar, Lisbon, 6. 7. October 2008; COST Action C25. addprint AG, Possendorf.

Martinez-Arguelles, G., Acosta, M.P., Dugarte, M., Fuentes, L., 2019. Life Cycle Assessment of Natural and Recycled Concrete Aggregate Production for Road Pavements Applications in the Northern Region of Colombia: Case Study. Transp. Res. Rec. J. Transp. Res. Board 2673, 397–406. https://doi.org/10.1177/0361198119839955

Masnadi, M.S., El-Houjeiri, H.M., Schunack, D., Li, Y., Englander, J.G., Badahdah, A., Monfort, J.-C., Anderson, J.E., Wallington, T.J., Bergerson, J.A., Gordon, D., Koomey, J., Przesmitzki, S., Azevedo, I.L., Bi, X.T., Duffy, J.E., Heath, G.A., Keoleian, G.A., McGlade, C., Meehan, D.N., Yeh, S., You, F., Wang, M., Brandt, A.R., 2018. Global carbon intensity of crude oil production. Science 361, 851–853. https://doi.org/10.1126/science.aar6859

Meili, C., Jungbluth, N., Annaheim, J., 2018. Life cycle inventories of crude oil extraction. https://doi.org/10.13140/RG.2.2.15479.27047

Miatto, A., Schandl, H., Fishman, T., Tanikawa, H., 2017. Global Patterns and Trends for Non-Metallic Minerals used for Construction: Global Non-Metallic Minerals Account. J. Ind. Ecol. 21, 924–937. https://doi.org/10.1111/jiec.12471

Mladenovič, A., Turk, J., Kovač, J., Mauko, A., Cotič, Z., 2015. Environmental evaluation of two scenarios for the selection of materials for asphalt wearing courses. J. Clean. Prod. 87, 683–691. https://doi.org/10.1016/j.jclepro.2014.10.013

Ohemeng, E.A., Ekolu, S.O., 2020. Comparative analysis on costs and benefits of producing natural and recycled concrete aggregates: A South African case study. Case Stud. Constr. Mater. 13, e00450. https://doi.org/10.1016/j.cscm.2020.e00450

Park, W.-J., Kim, T., Roh, S., Kim, R., 2019. Analysis of Life Cycle Environmental Impact of Recycled Aggregate. Appl. Sci. 9, 1021. https://doi.org/10.3390/app9051021

Pradhan, S., Tiwari, B.R., Kumar, S., Barai, S.V., 2019. Comparative LCA of recycled and natural aggregate concrete using Particle Packing Method and conventional method of design mix. J. Clean. Prod. 228, 679–691. https://doi.org/10.1016/j.jclepro.2019.04.328

Ritchie, H., Roser, M., 2022. Greenhouse gas emissions.

Rosado, L.P., Vitale, P., Penteado, C.S.G., Arena, U., 2017. Life cycle assessment of natural and mixed recycled aggregate production in Brazil. J. Clean. Prod. 151, 634–642. https://doi.org/10.1016/j.jclepro.2017.03.068

Saltelli, A., Annoni, P., 2010. How to avoid a perfunctory sensitivity analysis. Environ. Model. Softw. 25, 1508–1517. https://doi.org/10.1016/j.envsoft.2010.04.012

Silva, F., Cleto, F., Diestelkamp, E., Yoshida, O., de Oliveira, L., Saade, M., da Silva, V., Moraga, G., Passuello, A., Da Silva, M., Myers, N., Gmünder, S., 2018. Life Cycle Inventories of cement, Concrete and Related Industries -Brazil. SRI/Ecoinvent.

Simion, I.M., Fortuna, M.E., Bonoli, A., Gavrilescu, M., 2013. Comparing environmental impacts of natural inert and recycled construction and demolition waste processing using LCA. J. Environ. Eng. Landsc. Manag. 21, 273–287. https://doi.org/10.3846/16486897.2013.852558

Stadler, K., Wood, R., Bulavskaya, T., Södersten, C.-J., Simas, M., Schmidt, S., Usubiaga, A., Acosta-Fernández, J., Kuenen, J., Bruckner, M., Giljum, S., Lutter, S., Merciai, S., Schmidt, J.H., Theurl, M.C., Plutzar, C., Kastner, T., Eisenmenger, N., Erb, K.-H., de Koning, A., Tukker, A., 2018. EXIOBASE 3: Developing a Time Series of Detailed Environmentally Extended Multi-Regional Input-Output Tables: EXIOBASE 3. J. Ind. Ecol. 22, 502–515. https://doi.org/10.1111/jiec.12715

Stripple, H., 2001. Life cycle assessment of road. A Pilot Study for Inventory Analysis (No. Second Revised Edition). Report from the IVL Swedish EnvironmentalResearch Institute.

Tam, V.W.Y., 2008. Economic comparison of concrete recycling: A case study approach. Resour. Conserv. Recycl. 52, 821–828. https://doi.org/10.1016/j.resconrec.2007.12.001

Tošić, N., Marinković, S., Dašić, T., Stanić, M., 2015. Multicriteria optimization of natural and recycled aggregate concrete for structural use. J. Clean. Prod. 87, 766–776. https://doi.org/10.1016/j.jclepro.2014.10.070

WBCSD, WRI (Eds.), 2011. Product Life Cycle Accounting and Reporting Standard, Rev. ed. ed. World Business Council for Sustainable Development ; World Resources Institute, Geneva, Switzerland : Washington, DC.

Weidema, B.P., 1998. Multi-user test of the data quality matrix for product life cycle inventory data. Int. J. Life Cycle Assess. 3, 259–265. https://doi.org/10.1007/BF02979832

Weidema, B.P., Hischier, R., Mutel, C., Nemecek, T., Reinhard, J., Vadenbo, C.O., Wernet, G., 2013. Overview and methodology - Data quality guideline for the ecoinvent database version 3 (ecoinvent report No. No. 1). St. Gallen.

Yang, Y., Ingwersen, W.W., Hawkins, T.R., Srocka, M., Meyer, D.E., 2017. USEEIO: A new and transparent United States environmentally-extended input-output model. J. Clean. Prod. 158, 308–318. https://doi.org/10.1016/j.jclepro.2017.04.150

Zhang, Y., Luo, W., Wang, J., Wang, Y., Xu, Y., Xiao, J., 2019. A review of life cycle assessment of recycled aggregate concrete. Constr. Build. Mater. 209, 115–125. https://doi.org/10.1016/j.conbuildmat.2019.03.078

# Supplementary material - Understanding the environmental impacts of virgin aggregates: critical literature review and primary comprehensive Life Cycle Assessments

Anne de Bortoli[1,2,3,4]*

**1** CIRAIG, École Polytechnique de Montréal, P.O. Box 6079, Montréal, Québec, H3C 3A7, Canada

**2** Centre Technique Amériques, Eurovia Canada Inc., 3705 Place Java #210, Brossard, QC J4Y 0E4, Canada

**3** Direction technique, Eurovia Management, 18 Place de l'Europe, 92500, Rueil-Malmaison, France

**4** LVMT, Ecole des Ponts ParisTech, Cité Descartes, 6-8 Avenue Blaise Pascal, 77420 Champs-sur-Marne, France

\* Corresponding author; e-mail: anne.debortoli@polymtl.ca

# Table of content

# Table of tables

## Table of figures

# 1. Details on modeling of blasting

*Explosive consumption by type of rock*

Explosive activities are not accounted for in most aggregate LCAs, although a blasting process exists in the ecoinvent database. It includes the production of ammonium nitrate produced in Switzerland called tovex, its transport to the quarry, and airborne emissions based on stoichiometric calculations (EcoInvent, n.d.). According to data from Quebec sites (cf Denis Bouchard, Eurovia's expert), explosive consumption varies according to the type of rock: these consumptions per cubic meter of exploded rock are shown in Table 2.

The density of rocks varies according to their nature. The cellular rocks present densities around 2.55 t/m$^3$ while the rocks with metallic structures can go up to densities of 3.2 or 3.3 t/m$^3$, for chalcopyrite stones for example which contain iron sulfides. The rocks used for road aggregates in Quebec have densities in the range [2.55-2.85] $t/m^3$. I will consider an average density of 2.7 $t/m^3$ of rock to recalculate the consumption of explosives per ton. The results are presented in Table 2.

**Table 2 Consumption of explosives to blast different kinds of rocks**

| Type of rock | Explosive consumption | |
|---|---|---|
| | $kg/m^3$ of rock | kg/t of rock |
| Limestone rocks | 0.45 | 0.17 |
| Dolomitic rocks and slate | 0.70 | 0.26 |
| Hard rocks (volcanic, sandstone) | 1.00 | 0.37 |

*Explosive consumption on average and by installation type*

To create representative average explosive consumption models per type of Eurovia's Quebec production– for fixed and mobile installation, on average, in 2020 and 2021 – I calculated the product ratios and the associated rock types (Table 3).

**Table 3 Types of rocks extracted on average by Eurovia Quebec and by type of production unit**

| Type of rock | 2020 | 2021 |
|---|---|---|
| **Average production (t)** | | |
| Limestone rocks | 31% | 33% |
| Dolomitic rocks | 21% | 17% |
| Volcanic rocks | 46% | 48% |
| Loose rocks | 2% | 2% |
| **Total** | 100% | 100% |
| **Mobile production (t)** | **2020** | **2021** |
| Limestone rocks | 13% | 16% |
| Dolomitic rocks | 42% | 40% |
| Volcanic rocks | 45% | 43% |
| **Total** | 100% | 100% |
| | | |
| **Fixed production (t)** | **2020** | **2021** |
| Limestone rocks | 51% | 47% |
| Dolomitic rocks | 0% | 0% |
| Volcanic rocks | 49% | 53% |
| **Total** | 100% | 100% |

*Sums can differ from 100% due to round numbers*

Moreover, I calculated that exactly half of Eurovia's production came from fixed production units in 2020, and the rest from mobile units (resp. 58 and 42% in 2021), leading to the average consumption of explosives by type of production shown in the manuscript.

## 2. Details on modeling of energy and water consumption

The "mobile industry" virtual activity area corresponds to the production of mobile installations that operate on different sites which do not have their fixed crushing/screening unit. These sites consume diesel (10% of all the diesel consumed in Eurovia Quebec's production over the period), gasoline (1% of consumption), ORD (11%), LFO (48%), and electricity (29%). These services, therefore, represent significant consumption that I had to break down into each area of use. Additional site data were thus collected via Denis Bouchard, Vice-President of Quarries at Eurovia Québec, from its operations managers.

Moreover, these records must be corrected as they only account for direct consumptions, and not for sub-contracted consumptions. Thus, the records of site consumption tend to underestimate the real environmental impact of the production of Eurovia's aggregates. I was able to correct this by including a crushed tonnage consumption flat rate subcontracted by zone. The data considered was provided by Denis Bouchard and is presented in Table 15 and Table 16. They are rather lower than the values obtained for fixed productions via reporting data and would require detailed verification since the environmental impact of aggregates is strongly determined by the consumption of fossil fuels.

Table 15 Energy consumption for sub-contracted activities per ton of aggregate produced

| Energy | Consumption | Unit |
|---|---|---|
| Standard diesel | 0.22 | L/t |
| Off-road diesel | 0.12 | L/t |
| Electricity | 0.94 | kWh/t |

Table 16 Sub-contracted crushing production by zone

| REGION NAME | SUB-CONTRACTED CRUSHING (T) |
|---|---|
| GASPESIE QUARRIES | 0 |
| NORTH MATERIALS | 68 015 |
| AGGREGATES SAINTE-CLOTILDE | 799 242 |
| EASTERN TOWNSHIP QUARRIES | 90 042 |
| RSMM QUARRIES | 0 |
| OUTAOUAIS QUARRIES | 0 |

*Corrected energy and water consumption flows*

By dividing the consumption by the tonnage produced, I calculate the consumption per ton placed on the market. All data is shown in Table 17.

Table 17 Consumption flows per functional unit, production regions, and on Quebec average

| RÉGION | DIESEL (L/T) | GASOLINE (L/T) | LFO (L/T) | NATURAL GAS (M3/T) | ELECTRI-CITY (KWH/T) | WATER (M3/T) | ORD (L/T) | QUARRY TO CONSUMER DISTANCE (KM) |
|---|---|---|---|---|---|---|---|---|
| GASPESIE QUARRIES | 5.95E-01 | 1.27E-02 | 0.00E+00 | 0.00E+00 | 1.97E-01 | 0 | 3.09E-01 | 45.5 |
| NORTH MATERIALS | 5.12E-01 | 4.55E-03 | 5.16E-03 | 2.10E-02 | 4.11E+00 | 6.385E-05 | 4.49E-02 | 3.1 |
| AGGREGATES SAINTE-CLOTILDE | 7.80E-01 | 2.52E-03 | 1.03E-01 | 0.00E+00 | 1.76E+00 | 0 | 1.18E-01 | 32.9 |
| EASTERN TOWNSHIP QUARRIES | 3.63E-01 | 1.53E-02 | 2.45E-03 | 0.00E+00 | 7.48E-01 | 5.45E-07 | 1.32E-01 | 6.7 |
| RSMM QUARRIES | 4.00E-01 | 2.74E-02 | 1.42E-02 | 0.00E+00 | 2.47E+00 | 0 | 8.82E-02 | 19.9 |
| OUTAOUAIS QUARRIES | 5.38E-01 | 0.00E+00 | 0.00E+00 | 0.00E+00 | 1.47E-01 | 0 | 7.78E-02 | 7.9 |
| *QUEBEC AVERAGE* | *4.82E-01* | *1.42E-02* | *1.73E-02* | *4.25E-03* | *1.95E+00* | *1.31E-05* | *1.17E-01* | *16.9* |

## 3. Regionalization of GHGs from burned fuels

To account for the regional aspects of these consumptions, I use several regional data for Quebec and Canada. Only GHG emissions are regionalized, following the method and using the data detailed below. For instance, to model the combustion of ORD used by mobile machinery on quarries, I modify the ecoinvent process "diesel, burned in building machine, GLO". I readjust the mass consumption of fuel per MJ consumed based on LégisQuébec's lower calorific value (LCV) for diesel (38.3 MJ/l) (LégisQuébec, 2021) and a density of 0.85 kg/l ("The Engineering Toolbox," n.d.). I then use direct GHG emission factors from LégisQuébec. I then correct the other emissions by the factor 0.0222/0.0234 corresponding to the difference in PCI between LégisQuébec and ecoinvent. As no GHG emission factor is indicated on LégisQuébec for the ORD, I use those of the Ministry of the Environment (MELCC, 2019).

The other kinds of fuel used – gasoline, natural gas, LFO, and diesel for non-mobile machinery - will be rectified on the GHG they emit using the same methodology, using densities considered in Table 18, and the energy conversion and carbon emission factors shown in Table 19. These factors come from documents from the Quebec Ministry for the Energetic Transition (Quebec Ministry for the Energetic transition, 2019) and the document called "RDOCECA" from LégisQuébec (LégisQuébec, 2021). For densities, the "engineering toolbox" database is used when no terrain or Quebec data has been found ("The Engineering Toolbox," n.d.).

Table 18 Fuel densities

| FUEL | DENSITIES (KG/M3) | SOURCE |
|---|---|---|
| LFO | 860 | RDOCECA - table 1-10 |
| DIESEL | 850 | Engineering toolbox |

Table 19 Energy conversion and carbon emission factors

| NAME | UNIT | MJ/UNIT | GHG EMISSIONS (G) | | | | |
|---|---|---|---|---|---|---|---|
| | | | CO2 | CH4 | N2O | Source | Usage |
| REGULAR DIESEL, MOBILE | L | 38.3 | 2663 | 0.15 | 1.1 | LegisQuebec Table 27-1 | Mobile |
| ORD | L | 38.3 | 2681 | 0.073 | 0.022 | MELCC 2019 - Table 4 | Mobile, diesel HCV from LegisQuebec |
| REGULAR DIESEL, FIXED | L | 38.3 | 2663 | 0.133 | 0.4 | LegisQuebec Table 1-3 | Fixed |
| GASOLINE | L | 34.87 | 2289 | 2.7 | 0.05 | LegisQuebec Table 27-1 | Mobile, fixed HCV |
| NON ROAD GASOLINE | L | | 2037 | 19.61 | 0.013 | MELCC 2019 - Table 4 | |
| NATURAL GAS | m3 | 37.89 | 1878 | 0.037 | 0.034 | FEC Quebec | Mix of the HCV of natural gas & distillation gas, fixed, CO2/m3 table 1-4 LegisQuebec |
| NATURAL GAS | MJ | 1 | 50.7 | $9.99\text{E}-4$ | $9.18\text{E}-4$ | FEC Quebec | |
| LFO (OIL #2) | L | 38.5 | 2725 | 0.006 | 0.031 | LegisQuebec Table 1-3 | Fixed |
| LFO (OIL #2) | MJ | 1 | 70.8 | $1.56\text{E}-4$ | $8.06\text{E}-4$ | FEC Quebec | |

## 4. Details on dust emissions

Dust emissions are not measured on-site in Quebec. Nevertheless, quarries producing more than 500,000 t/year of aggregates must report their emissions in the Canadian National Pollution Inventory (NPRI). Default emission factors are recommended by the Government of Canada depending on the type and method of production. These data come from the 5th report of the USA EPA, chapter 11, corresponding to the environmental impacts of the mineral industry (US EPA, 2004).

## 5. Site characteristics

**Table 20 Characteristics of three fixed unit facilities**

| Site | Type | Fixed facilities | | | Mobile unit | | Unit |
|---|---|---|---|---|---|---|---|
| | Name | LAVAL | Saint-Bruno | Saint-Philippe | SDE | SBE | |
| | Annual production | 1200 | 1100 | 775 | 430 | 430 | kt |
| | Site lifespan | 40 | 30 | 65 | 25 | 25 | years |
| | Site production | 4.80E+10 | 3.30E+10 | 5.04E+10 | 1.08E+10 | 1.08E+10 | kg |
| Lad occupation | Waterbody | 55 000 | 2875 | 40000 | X | X | m² |
| | Rest | 1 815 000 | 525000 | 830000 | X | X | m² |
| Concrete elements | Primary slab | 750 | 225 | 60 | X | X | m3 |
| | Primary electric sub-station | 14 | 135 | 75 | X | X | m3 |
| | Secondary slab | 900 | 16 | 75 | X | X | m3 |
| | Scalper and tower | 46.5 | 180 | 16 | X | X | m3 |
| | Third slab | 540 | 668 | 578 | X | X | m3 |
| | Main sub-station | 17 | 13 | 180 | X | X | m3 |
| | Pumping station | X | 6 | X | X | X | m3 |
| | Total | 2268 | | 984 | X | X | m3 |
| Conveyor belt | Length | 1680 | 1705 | 1000 | X | X | m |

## 6. Capital goods

The basin is dug into the rock and does not include any particular structure (neither concrete nor liner). Concrete infrastructure was built to accommodate the various elements of the crushing unit. The different slabs are detailed in an Excel spreadsheet, and the total masses of 30-32 MPa concrete used have been compiled in the supplementary material. The masses of industrial machinery and metal structures were roughly assessed on the metal frame part because no traceability document allowed us to make a more precise estimate. The quarries also have roads: they are often paved for customer access near cities (small sites) but remain unpaved on larger production sites. They will therefore be neglected as most of the sites are unpaved. The data relating to the equipment outside the conveyor belt could only be collected on the Laval quarry and are presented in the supplementary material, for a lifespan of 25 years, and lead to amortizing 10.7 g of industrial machines per ton of aggregates produced. We also need to assess LULUC. The lands occupied were previously generally woods or agricultural fields, estimated under a ratio of resp. 30-70% in Quebec. The redevelopment is very variable, i.e., the working faces can be left as they are or further redeveloped.

Table 21 Mobile machinery amortization

| Component | Unitary weight (kg) |
|---|---|
| Primary crusher | 26000 |
| Mobile sieve | 13500 |
| Secondary crusher | 24000 |
| Double sieve | 21000 |
| Third crusher | 26000 |
| Container | 26000 |
| Trailer | 18000 |
| Total (kg) | 154500 |
| Machinery amortization (kg/kg of rock) | 1.44E-05 |

Unit_mass / (Annual production * Lifespan) = 14.4 g

Table 22 Fixed machinery amortization

| COMPONENT | NUMBER | UNITARY WEIGHT (LBS) | UNITARY WEIGHT (KG) |
|---|---|---|---|
| PRIMARY CRUSHER | 1 | | 72575 |
| SCALPER SIEVE | 1 | 24500 | 24500 |
| SECONDARY CRUSHER | 3 | | 80592 |
| SECONDARY SIEVES | 3 | 14500 | 43500 |
| THIRD CRUSHER | 1 | | |
| THIRD SIEVE | 2 | 17500 | 35000 |
| THIRD SIEVE | 2 | 19800 | 39600 |
| PRIMARY SYNTRON | 2 | 4600 | 9200 |
| SECONDARY SYNTRON | 3 | 5300 | 15900 |
| | | TOTAL (KG) | 320867 |
| | | MACHINES AMORTIZATION (T/KG) | 1.07E-08 |

The different spare parts found in a crushing unit are the feeders that feed the stones in the crushing unit, the sleeves that channel the stones between the different crushers and the liners that carry the stones in each crusher, the hammer and shield/mantle that are respectively the moving and fixed parts of the crushers for the fixed/mobile crushing units. Finally pads are metallic parts placed between the different parts.

## 7. Downstream transportation modeling

### 1. *System boundaries*

The truck transportation processes include the manufacturing of the truck, their maintenance, as well as their use stage. The lifespan of the truck is considered to be 1 000 000 km long. The amortization of the road is not accounted for.

### 2. *Average loads and distances*

Upstream transport - between the pit and the processing facility - is considered in the site energy consumption data presented above. Data on the downstream transport of aggregates were also provided by Denis Bouchard, vice-president of materials at Eurovia Quebec: average supply distances from quarries to customers, type of truck used, and average full load. The information is summarized in Table 7. In addition, no double freight can generally be set up. Thus, the new aggregates transport models will consider trucks empty returns. The average transport distance for aggregates is 16.9 km between quarries and Eurovia's customers in Quebec. Average full loads vary by the quarry, with a 2020 production weighted average of 32.8t, but an average of 35.5 and 22.0 tons resp. for semi-trailers and 12-wheel trucks.

**Table 7 Transportation characteristics for aggregates, from the quarry to the consumer, by region**

| Production zone | Vehicle type | Average distance from quarry to consumer (km) | Average full load (t) |
|---|---|---|---|
| GASPESIE QUARRIES | Semi-trailer | 45.5 | 35 |
| NORTH MATERIALS | 12-wheel trucks & semi-trailer | 3.1 | 22 |
| AGGREGATES SAINTE-CLOTILDE | Semi-trailer | 32.9 | 36 |
| EASTERN TOWNSHIP QUARRIES | Semi-trailer | 6.7 | 34 |
| RSMM QUARRIES | Semi-trailer | 19.9 | 37 |
| OUTAOUAIS QUARRIES | Semi-trailer | 7.9 | 34 |
| QUEBEC WEIGHTED AVERAGE | | *16.9* | *35.5* |

## 3. *Fuel consumption*

The fuel consumption for the trucks used by Eurovia in Quebec is shown in Table 8. These data relate to average driving cycles for the semi-trailer truck, in full-load or empty modes, while rural and urban modes have been differentiated for the 12-wheel truck, but not the load condition. Consumption is higher than the average for Quebec's truck fleet, evaluated at 39.5L/100km by a provincial study carried out in 2000 (Government of Canada, 2009). This can be explained by the high tonnages of materials transported by the trucks used in construction. Moreover, the Quebec survey refers to a consumption difference of 5L/100km between summer and winter. I assumed that the data collected from Eurovia rather relate to summer consumption, as road construction activities mostly stop in winter in Quebec due to the cold weather.

**Table 8 Fuel consumption of trucks depending on the load and driving conditions**

| Vehicle | Load & driving conditions | Consumption |
|---|---|---|
| Semi-trailer truck | Full, average | 50L/100km |
|  | Empty, average | 39L/100km |
| 12-wheel truck | Average, rural | 40L/100km |
|  | Average, urban | 52L/100km |

*4.  Emissions*

Only emissions from small trucks (under 4.54 t) are regulated in Canada, based on US EPA regulations. The reference emission models for vehicles in North America are those of the MOVES models, and particularly of MOVES3 which updated the data dating from 2009 in the previous version of MOVES (US EPA, 2020). But I did not find a way to extract the unitary emissions of the vehicles from MOVES3. Thus, I chose to use the European vehicle emission models from HBEFA v4.1 (Matzer et al., 2019). Despite the geographical difference, HBEFA has the advantage of simulating consumption and emissions in real conditions. I considered the average Swiss truck fleet in 2020 and calculated the fuel consumption (FC) as well as the regulated emissions (CO, $CO_2$, NOx, HC, PM) of heavy trucks.

**Table 23 Fuel consumption (FC) and polluting emissions due to truck exhaust fumes**

| Country | Year | Vehicle category | Flow | Emission factor | Unit |
|---|---|---|---|---|---|
| CH | 2020 | HGV | CO | 0.742 | [g/Vehkm] |
| CH | 2020 | HGV | CO2 | 833.324 | [g/Vehkm] |
| CH | 2020 | HGV | FC | 264.538 | [g/Vehkm] |
| CH | 2020 | HGV | HC | 0.075 | [g/Vehkm] |
| CH | 2020 | HGV | NOx | 1.764 | [g/Vehkm] |
| CH | 2020 | HGV | PM | 0.03 | [g/Vehkm] |

However, the simulated consumption does not correspond to heavily loaded large-capacity trucks used in construction. Thus, I adjusted the emissions to the consumption data of Eurovia's fleet, considering consumption and emissions to be proportional to HBEFA results above. Final fuel consumption and emissions of trucks to transport aggregates in different operating conditions are presented in Table 9.

**Table 9 Consumptions and emissions of aggregate trucks in different operating conditions (per 100 km)**

| | | Conso (L/100km) | Ratio HBEFA/Eurovia | CO | CO2 | HC | NOx | PM |
|---|---|---|---|---|---|---|---|---|
| Semi-trailer truck | Full, average | 50 | 1.61 | 1.19E+00 | 1.34E+03 | 1.20E-01 | 2.83E+00 | 4.82E-02 |
| | Empty, average | 39 | 1.25 | 9.30E-01 | 1.04E+03 | 9.40E-02 | 2.21E+00 | 3.76E-02 |
| | Double trip (empty return) | 89 | 2.86 | 2.12E+0 | 2.38E+3 | 2.10E-1 | 5.04E+0 | 8.60E-2 |
| 12-wheel truck | Rural, average | 40 | 1.29 | 9.54E-01 | 1.07E+03 | 9.64E-02 | 2.27E+00 | 3.86E-02 |
| | Urban, average | 52 | 1.67 | 1.24E+00 | 1.39E+03 | 1.25E-01 | 2.95E+00 | 5.01E-02 |
| | Double trip (empty return) | 92 | 2.96 | 2.19E+0 | 2.46E+3 | 0.22E-1 | 5.21E+0 | 8.90E-2 |

Other exhaust emissions (CH4, N2O, S2O) in the processes created are based on the truck transportation process developed in the US LCI database.

## 5. *Comparison of transportation processes' impacts*

The load being much higher in the transport of construction materials than in that of the average truck (5.79 t in ecoinvent for the 16-32 t category, 15.96 t for the <32 t category), despite taking into account empty returns, the environmental impacts are much less important with my new models than with ecoinvent's default truck transport model (in black on the histogram below).

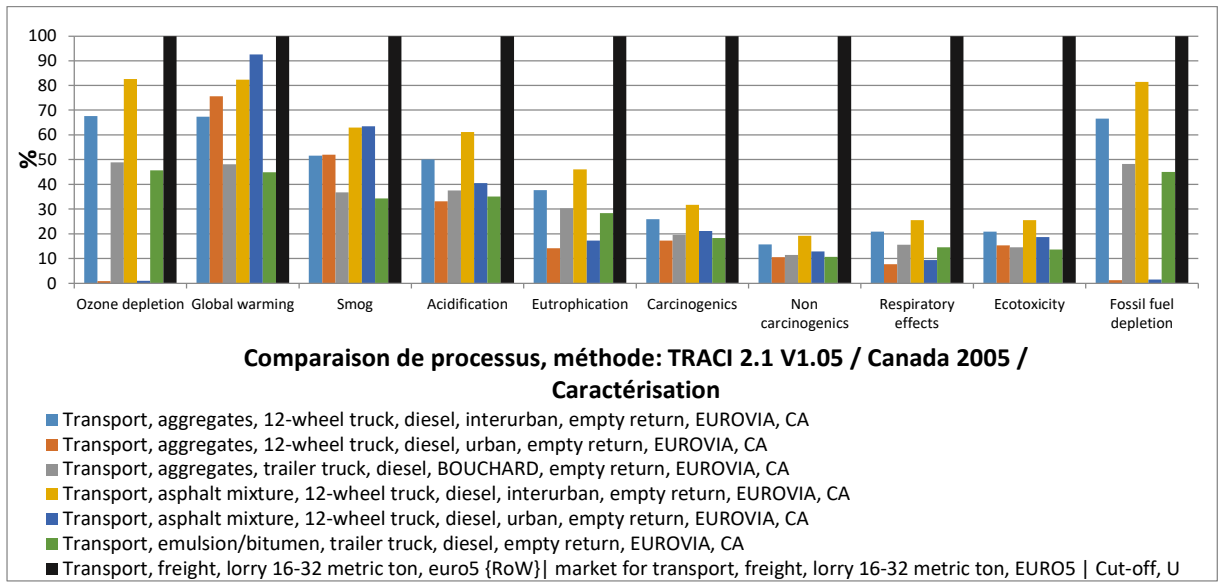

**Figure 10 Comparison of transportation processes' impacts**

A focus on GHG emissions is presented in the histogram below: it highlights, in particular, the low carbon footprint of transport per tkm for construction: between 77 gCO2eq/tkm for the transport of binders and 159 gCO2eq/tkm for asphalt transported in 12-wheel trucks in urban areas, compared to a default value in ecoinvent of 172 gCO2eq/t for 16-32t ecoinvent truck category and 0.090 gCO2eq/t for >32t truck category.

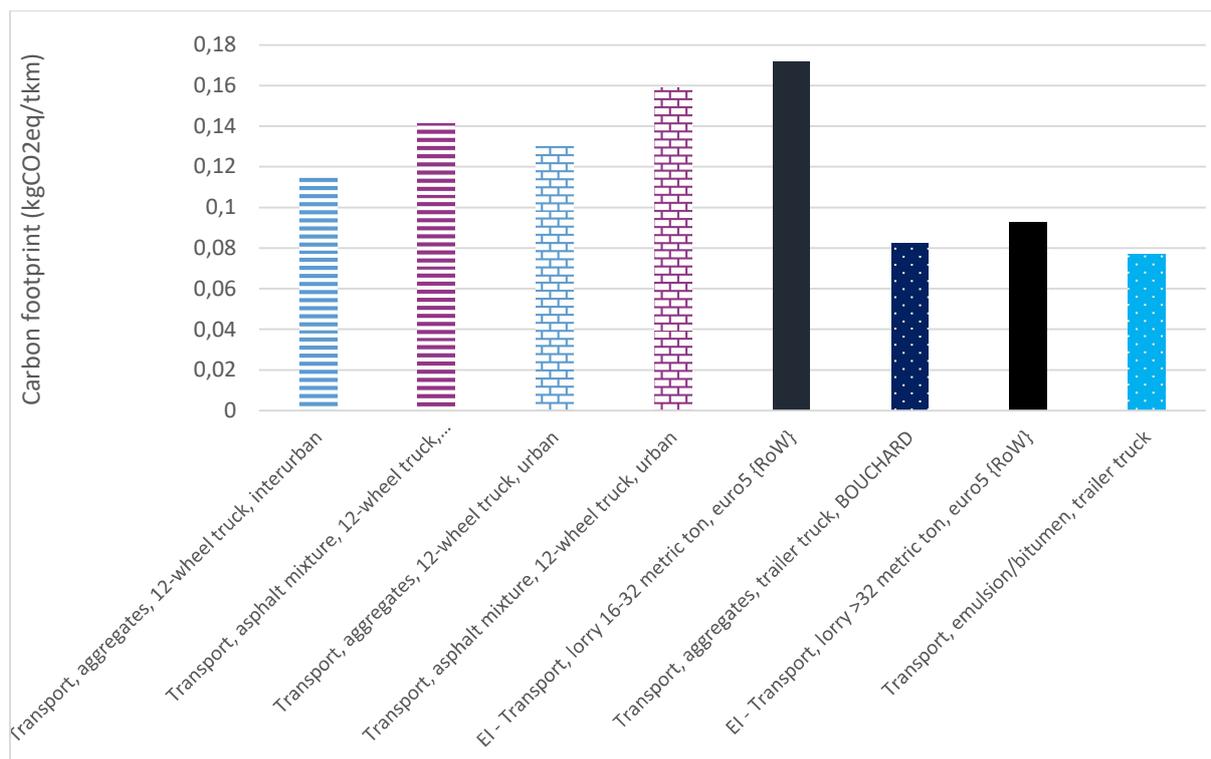

Figure 11 4.    Comparison of transportation processes carbon footprints

# 8. Results of the scenario analysis on electricity mixes

Table 24 Impact of aggregates depending on the electricity mix

| Impact category | Unit | **Canada-Alberta** | **Canada-average** | **Canada-Ontario** | **Canada-Quebec** | **China** | **USA - Midwest** | **USA - Northeast** | **Europe** | **USA - Western** |
|---|---|---|---|---|---|---|---|---|---|---|
| Ozone depletion | kg CFC-11 eq | 5.25E-07 | 4.95E-07 | 4.95E-07 | 4.76E-07 | 4.91E-07 | 5.22E-07 | 5.95E-07 | 5.76E-07 | 5.42E-07 |
| Global warming | kg CO2 eq | 4.98E+00 | 3.66E+00 | 3.40E+00 | 3.28E+00 | 5.15E+00 | 4.55E+00 | 3.70E+00 | 4.04E+00 | 4.10E+00 |
| Smog | kg O3 eq | 2.90E+00 | 2.85E+00 | 2.84E+00 | 2.84E+00 | 3.01E+00 | 2.89E+00 | 2.85E+00 | 2.87E+00 | 2.87E+00 |
| Acidification | kg SO2 eq | 1.24E-01 | 1.19E-01 | 1.18E-01 | 1.18E-01 | 1.36E-01 | 1.23E-01 | 1.19E-01 | 1.21E-01 | 1.20E-01 |
| Eutrophication | kg N eq | 3.53E-02 | 1.55E-02 | 1.15E-02 | 1.14E-02 | 1.42E-02 | 2.77E-02 | 1.18E-02 | 1.74E-02 | 1.85E-02 |
| Carcinogenics | CTUh | 6.85E-07 | 5.05E-07 | 4.70E-07 | 4.63E-07 | 5.34E-07 | 6.16E-07 | 4.71E-07 | 5.27E-07 | 5.36E-07 |

| Non carcinogenics | CTUh | 1.47E-06 | 8.88E-07 | 7.94E-07 | 7.40E-07 | 1.01E-06 | 1.33E-06 | 7.96E-07 | 9.92E-07 | 9.86E-07 |
| --- | --- | --- | --- | --- | --- | --- | --- | --- | --- | --- |
| Respiratory effects | kg PM2.5 eq | 8.28E-03 | 8.08E-03 | 7.93E-03 | 7.87E-03 | 9.91E-03 | 1.47E-02 | 8.00E-03 | 8.37E-03 | 1.07E-02 |
| Ecotoxicity | CTUe | 7.85E+01 | 6.06E+01 | 5.85E+01 | 5.56E+01 | 6.25E+01 | 7.28E+01 | 5.78E+01 | 6.33E+01 | 6.39E+01 |
| Fossil fuel depletion | MJ surplus | 5.47E+00 | 4.83E+00 | 4.82E+00 | 4.52E+00 | 4.82E+00 | 4.86E+00 | 5.52E+00 | 5.13E+00 | 5.30E+00 |

# 9. Bibliography

LégisQuébec, 2021. Q-2, r. 15 - Règlement sur la déclaration obligatoire de certaines émissions de contaminants dans l'atmosphère, Loi sur la qualité de l'environnement.

MELCC, 2019. Guide de quantification des émissions de gaz à effet de serre. Ministère de l'Environnement et de la lutte contre les changements climatiques.

Quebec Ministry for the Energetic transition, 2019. Facteurs d'émission et de conversion.

The Engineering Toolbox [WWW Document], n.d. URL https://www.engineeringtoolbox.com/material-properties-t_24.html